\documentclass[preprint2]{aastex}

\renewcommand{\deg}{\mbox{$^{\circ}$}}
\newcommand{\kms}{\mbox{km s$^{-1}$}}

\newcommand{\uJy}{\mbox{$\mu$Jy}}
\newcommand{\um}{\mbox{$\mu$m}}

\newcommand{\x}{\mbox{$\times$}}
\newcommand{\Msun}{\mbox{$M_{\odot}$}}

\newcommand{\Lsun}{\mbox{$L_{\odot}$}}
\newcommand{\E}[1]{\hbox{$10^{ #1 }$}}
\newcommand{\per}[1]{\hbox{#1$^{-1}$}}

\newcommand{\about}{\mbox{$\sim$}}

\def\IRAS       {\hbox{\it IRAS}}
\def\NVSS       {\hbox{\it NVSS}}

\def\NO         {\hbox{---}}

\begin{document}


\title{COLA III. Radio Detection of AGN in Compact Moderate Luminosity Infra-Red Galaxies}

\author{R.~Parra\altaffilmark{1}\altaffiltext{1}{Now at European
    Southern Observatory, Alonso de Cordova 3107, Casilla 19001,
    Santiago 19, Chile}, J.~E. Conway, S.~Aalto } \affil{Onsala Space
  Observatory, SE-439 92 Onsala, Sweden}

\author{P.~N. Appleton} \affil{NASA Herschel Science Center, Mail Code
  100-22, California Institute of Technology. 770 S. Wilson
  Blvd. Pasadena CA 91125}

\author{R.~P. Norris} \affil{CSIRO Australia Telescope National
  Facility, P.O. Box 76, Epping, NSW 17 10, Australia}

\author{Y.~M. Pihlstr\"om\altaffilmark{2}\altaffiltext{2}{Also Adjunct
    Astronomer at the National Radio Astronomy Observatory}}
\affil{Department of Physics and Astronomy, UNM, 800 Yale Blvd NE,
  Albuquerque, NM 87131}

\and \author{L.~J. Kewley} \affil{University of Hawaii, 2680 Woodlawn
  Drive, Honolulu, HI 96822}

\begin{abstract}
  We present results from 4.8\,GHz VLA and Global-VLBI observations of
  the northern half of the moderate FIR luminosity (median $L_{\rm IR}
  = \E{11.01}\Lsun$) COLA sample of star-forming galaxies. VLBI
  sources are detected in a high fraction (20/90) of the galaxies
  observed. The radio luminosities of these cores
  ($\about$\E{21}\,W\,\per{Hz}) are too large to be explained by radio
  supernovae or supernova remnants and we argue that they are instead
  powered by AGN. These sub-parsec scale radio cores are
  preferentially detected toward galaxies whose VLA maps show bright
  100-500\,parsec scale nuclear radio components. Since these latter
  structures tightly follow the FIR to radio-continuum correlation for
  star-formation we conclude that the AGN powered VLBI sources are
  associated with compact nuclear starburst environments. The
  implications for possible starburst-AGN connections are discussed.
  The detected VLBI sources have a relatively narrow range of radio
  luminosity consistent with models in which intense compact
  Eddington-limited starbursts regulate the gas supply onto a central
  super-massive black hole. The high incidence of AGN radio cores in
  compact starbursts suggests little or no delay between the starburst
  phase and the onset of AGN activity.
  
\end{abstract}
\keywords{galaxies: active --- galaxies: starburst --- infrared:
  galaxies --- radio continuum: galaxies}
\maketitle

\section{Introduction}\label{se:INTRO}

It seems likely that super-massive black holes (SMBH) are ubiquitous
in galactic nuclei. The nuclear stellar-velocity dispersion and
bulge-size to black hole mass correlations
\citep{MAGORRIAN98,MERLONI03,HARING04} indicate that most galaxies
with central bulges should contain SMBHs at their centers. The
presence of a SMBH is a necessary condition for AGN activity but only
a small fraction of galaxies at any given time are strong AGN sources,
implying that a decisive factor determining AGN luminosity is the gas
feeding mechanism. While for Seyfert or LINER class AGN only
relatively small black hole accretion rates are required (0.01 to
0.1\,\Msun\,\per{year}) gas originating in the galactic disk must lose
angular momentum in order to be accreted onto the central
black hole. Various mechanisms to achieve such momentum transfer and
mass transport have been suggested \citep[see review by][]{KNAPEN04}.

There is increasing observational evidence for an AGN-starburst
connection \citep{DAVIES07, HECKMAN08, PRIETO05,PRIETO07,FATHI06}.
Plausibly such circumnuclear starbursts may play a role in removing
angular momentum and so feeding the central black hole
\citep{SCHART10}. Alternatively both nuclear starbursts and AGN could
be independent consequences of a common gas transport mechanism into
the centers of galaxies with no direct causal connection between the
two phenomena. Understanding in detail the phenomenology of starburst
and AGN activity, in particular whether one phase begins before the
other, is important to test the nature of the starburst-AGN
inter-relationship.

Most detailed observational work so far on the starburst-AGN
connection has studied starburst activity in samples of AGN
\citep{DAVIES07}. Here we adopt the alternative approach of searching
for AGN activity in a far infrared-selected sample of star-forming
galaxies known as COLA \citep[Compact Objects in Low-power AGN,][see
\S2 for sample definition]{CORBETT02,CORBETT03}. Our principal means
of identifying AGN activity in this sample is by searching for high
brightness temperature VLBI radio cores. Classic optical emission line
diagnostics \citep{BPT81,VO87,KEWLEY01,KAUF03} can be unreliable for
such searches as an AGN may be obscured by dust in star-forming
galaxies.  High brightness-temperature radio emission is unambiguous
evidence of an AGN, however a non-detection does not exclude an AGN
because of the existence of both radio loud and radio quiet objects
\citep[see][for a recent review]{ZAMFIR08}; it follows that VLBI
searches can only give a lower limit to the AGN fraction. It should be
noted that while radio supernovae (SNe) can also generate compact
radio emission they have a maximum luminosity which distinguishes them
from AGN \citep[see][and \S\ref{SNe}]{CORBETT02}.

We have already published radio observations \citep{CORBETT02} and
optical spectroscopy results \citet{CORBETT03} for the $\delta < 0
\deg$ (COLA-S) part of the sample. Here we present radio and optical
observations of the Northern half ($\delta > 0 \deg$) of the sample
(COLA-N).  This paper is organized as follows. In \S\ref{se:COLA} we
give an overview of the COLA project and previous observations. In \S3
we describe our VLA, VLBI, optical spectroscopy and other observations
and give an overview of the data reduction process. Our observational
results are presented in \S4 and our discussion in \S5. Finally in \S6
we list our major conclusions.

\section{The COLA Project and Previous Observations}
\label{se:COLA}
The COLA (Compact Objects in Low-power AGN) project has the primary
goal of determining the relationship between AGN and other galactic
properties (such as galactic structure, degree of interaction and star
formation activity) in a moderate luminosity FIR selected sample. The
sample contains all the galaxies in the \IRAS\ Point Source Catalog
\citep{IRAS88} with (a) flux densities at 60\,\um\ greater than 4\,Jy,
(b) heliocentric velocities between 3500 and 7000~\kms\ (to minimize
Malmquist bias) and (c) galactic latitude $|b|$$>$10$^{\circ}$ (to
avoid confusion with galactic objects).  The resulting sample has 217
galaxies of which 110 are located in the northern hemisphere (COLA-N)
and 107 in the Southern hemisphere (COLA-S).

The COLA sample is selected at the same IR wavelength as the Revised
Bright Galaxy Sample (RBGS see \citet{SANDERS03}) but with a flux
limit which is somewhat smaller (4\,Jy versus 5.24\,Jy) plus an
additional redshift range selection. It follows that there is a
significant overlap between the two samples with 65/110 (59\%) of the
COLA-N sources also being members of the RBGS.

Table\,1 provides a summary of the COLA-N sample \citep[for a similar
table describing the southern sample see][]{CORBETT02}.  The
luminosity distance $D_{L}$ given column~7 for each source was
calculated in NED assuming the three attractor model of
\citet{MOULD00} and the cosmological parameters derived from 5 year
WMAP data \citep{HINSHAW09}, i.e.  $\Omega_{m}=0.276$,
$\Omega_{v}=0.726$ and $H_{o} = 70.5$ kms$^{-1}$. Details on the
calculation of Infra-red and radio luminosities presented in Table\,1
are given in \S\ref{se:archival}. Table\,1 shows that most COLA
sources have bolometric IR luminosities close to the boundary defining
Luminous Infra-Red galaxies (at $L_{\rm IR}$=\E{11}$\Lsun$,
\cite{SANDERS96}) with a median IR bolometric luminosity for COLA-N of
$\E{11.01}\Lsun$. All but one source has a bolometric IR luminosity in
the range $\E{10.5} <L_{\rm IR}/\Lsun<$\E{11.7}, the exception being
the nearby ULIRG Arp\,220.

Low resolution 1.4, 2.5 and 4.8\,GHz continuum observations of COLA-S
sources have been made using the Australia Telescope Compact Array
(ATCA). These radio observations were complemented with high
resolution single baseline 2.3\,GHz snapshot observations obtained
using the Australian Long Baseline Array (LBA). Optical spectroscopy
for a large fraction of the southern sources was also obtained using
the Dual Beam Spectrograph (DBS) on the 2.3\,m telescope at the Mount
Stromlo and Siding Springs Observatory.  Detailed descriptions of the
COLA-S radio and optical observations can be found in
\citet{CORBETT02,CORBETT03} respectively.

\citet{CORBETT02} detected 9 out of the 105 COLA-S galaxies in their
LBA observations. Of these, 8 showed radio emission stronger than that
predicted by the IR-radio correlation. Moreover 7 of the 14 sources
with a total radio flux density more than 1$\sigma$ above that
predicted by the FIR-Radio correlation were LBA detections. Based on
these results \citet{CORBETT02} concluded that sources detected with
the LBA exhibited a statistically significant radio excess relative to
the non-detections. This excess persisted even after the subtraction
of the radio emission from the detected core implying the existence of
associated AGN powered diffuse radio emission from structures larger
than \about15\,pc.

The COLA-S optical data \citep{CORBETT03} showed that in the southern
sample the fraction of galaxies optically classified as Seyfert was
\about15\%. Of these only 55\% were detected with the LBA suggesting
the there may be two populations of Seyferts, one population with
extended radio structures and compact radio cores and the other
without. In this paper we revisit these tentative conclusions in light
of the larger number of long-baseline detected sources in COLA-N.

\section{Observations}
\subsection{VLA Observations}\label{se:VLA}
Snapshot radio continuum observations for the COLA-N sources were
obtained using the VLA at 4.8\,GHz in two observing epochs each of
24\,hours in length (see Table\,\ref{fi:vla_maps_detections}). Sources
with $0^{\circ}<\delta <30^{\circ}$ were observed on 1998/06/23 in the
VLA BnA configuration, while sources with $\delta>30^{\circ}$ were
observed on 2003/07/19 in the A configuration. In addition, to ensure
that all compact sources were observed with the A-array, 12 sources
found to be compact at 4.8\,GHz from the first epoch were re-observed
in the second epoch.  The data were reduced using AIPS in a standard
manner. Radio maps of size 50\arcsec\x50\arcsec\ centered at the
\IRAS\ positions were produced for all the sources using natural
weighting. Typical FWHM beam-sizes at 4.8~GHz were \about 0.4\arcsec\
and 0.9\arcsec\x0.4\arcsec\ for A and BnA configurations respectively.
The typical rms noise levels were 75\,\uJy\,\per{beam}

\subsection{VLBI Observations}\label{se:EVN}
VLBI 4.8\,GHz snapshot observations at data-rate 1~Gbit~s$^{-1}$ per
station were conducted for the COLA North sample using the most
sensitive telescopes available in the European VLBI Network (EVN). The
first epoch on Feb 27/28th 2005 used Eb (Effelsberg), Wb (Westerbork)
and Ar (Arecibo) for all sample sources within the declination range
$10^{\circ}<\delta <40^{\circ}$ (defined by the Ar zenith-distance
restrictions and minimum elevation at the European stations) with
\about15\,minutes of on-source time per target giving
$\sigma$=25\,$\mu$Jy on the Ar--Eb baseline. The second epoch on May
6/7th 2005 used Eb, Wb and Jb1 (Jodrell Bank, Lovell telescope) for
all the sources with $0^{\circ}<\delta <10^{\circ}$ and $\delta
>40^{\circ}$ with \about15\,minutes of on-source time per target
giving noise $\sigma=$56\,$\mu$Jy on the Eb--Jb1 baseline. Gaps in the
schedule were filled by re-observing 21 sources from the first
epoch. In both epochs strong nearby phase reference and amplitude
calibrators ($>$500\,mJy) were observed for 10\,minutes every
\about1\,hour. The Eb--Wb baseline (266\,km) which is common to both
epochs has a fringe spacing of 46.5\,mas at 4.8\,GHz which is similar
to the 47.4\,mas fringe-spacing of the ATCA--Tidbinbilla baseline
(566~km) at 2.3\,GHz used by \citet{CORBETT02} to observe the southern
sample.  Because of scheduling restrictions and bad data we only
obtained useful VLBI observations for 90 sources.

AIPS was used for the initial data reduction and calibration stages
with instrumental phase and delay being determined toward strong
calibrators.  Monitored system temperatures and gains were used to
calibrate amplitudes.  An adjustment of \about30\% was required on Jb1
to obtain consistent calibration on sources known to be unresolved
from the VLBA calibrator database. Our final estimate is that quoted
baseline flux densities are accurate to \about15\%.

Fringe fitting was performed in AIPS on all the calibrators and then
these solutions were applied to the interleaved target source
scans. After this we exported the data to our own software within
which Delay--Rate maps were made for each baseline/target source. The
size of these delay-rate maps were set such that a compact source
would be detectable up to 20\,\arcsec\ from the correlation
position. Finally the Delay--Rate maps were transformed into RA--Dec
offset maps relative to the phase center. These maps show \emph{only}
features which are smaller than the baseline fringe spacing
\citep[see][]{THESIS}. We adopted a detection threshold of 6 times the
measured r.m.s. noise. The high sensitivity and small fringe spacing
of the Eb--Ar baseline resulted in a very small uncertainty on the
phase center offsets of the detections. These VLBI position offsets
are shown as crosses on the corresponding VLA map in
Figure~\ref{fi:vla_maps_detections}


\begin{deluxetable}{ c c c c c c c c c c }
\centering
\tablecolumns{10}
\tablenum{1}
\tabletypesize{\scriptsize}
\tablewidth{0pt}
\tablecaption{\sc COLA North Sample}
\tablehead{
\colhead{IRAS}&
\colhead{Other}&
\colhead{R.A.}&
\colhead{Decl.}&
\colhead{$V_{\rm Hel}$}&
\colhead{$D_{L}$} & 
\colhead{$\log L_{\rm IR}$}&
\colhead{$\log L_{\rm 1.4}$}&
\colhead{$\log L_{\rm FIR}$}&
\colhead{$q_{1.4}$}
\\
\colhead{Name}&
\colhead{Name}&
\colhead{J2000}&
\colhead{J2000}&
\colhead{[\kms]}&
\colhead{[Mpc]}&
\colhead{[\Lsun]}&
\colhead{[W~\per{Hz}]}&
\colhead{[W~\per{Hz}]}&
\colhead{}
\\
\colhead{(1)}&
\colhead{(2)}&
\colhead{(3)}&
\colhead{(4)}&
\colhead{(5)}&
\colhead{(6)}&
\colhead{(7)}&
\colhead{(8)}&
\colhead{(9)}&
\colhead{(10)}
}
\startdata
00005+2140  	& MRK334  	& 00 03 09.74 	& 21 57 36.46 & 6579  	& 93.4 	& 11.07 	& 22.46 	& 24.74 	& 2.27 \\ 
00073+2538  	& N23  	& 00 09 53.58 	& 25 55 26.40 & 4566  	& 64.7 	& 11.09 	& 22.57 	& 24.82 	& 2.25 \\ 
00506+7248  	&   	& 00 54 04.00 	& 73 05 11.70 & 4706  	& 69.3 	& 11.49 	& 22.82 	& 25.21 	& 2.40 \\ 
00521+2858  	&   	& 00 54 50.27 	& 29 14 47.60 & 4629  	& 65.2 	& 10.88 	& 22.39 	& 24.62 	& 2.23 \\ 
00548+4331  	& N317  	& 00 57 40.58 	& 43 47 32.30 & 5429  	& 77.2 	& 11.16 	& 22.66 	& 24.92 	& 2.26 \\ 
00555+7614  	&   	& \emph{00 59 15} 	& \emph{76 30 52} & 4739  	& 70.1 	& 10.91 	& 22.53 	& 24.66 	& 2.13 \\ 
01503+1227  	&   	& 01 52 59.52 	& 12 42 27.90 & 4558  	& 63.3 	& 10.91 	& 22.36 	& 24.65 	& 2.29 \\ 
01519+3640  	&   	& 01 54 53.93 	& 36 55 04.40 & 5621  	& 79.2 	& 11.02 	& 22.14 	& 24.74 	& 2.60 \\ 
01555+0250  	& ARP126  	& 01 58 05.26 	& 03 05 00.90 & 5431  	& 75.6 	& 10.94 	& 22.47 	& 24.68 	& 2.20 \\ 
01556+2507  	&   	& 01 58 30.63 	& 25 21 36.90 & 4916  	& 68.8 	& 10.99 	& 22.36 	& 24.75 	& 2.39 \\ 
01579+5015  	&   	& 02 01 09.65 	& 50 30 25.50 & 4875  	& 69.5 	& 10.85 	& 22.29 	& 24.55 	& 2.26 \\ 
02071+3857  	& N828  	& 02 10 09.52 	& 39 11 25.30 & 5374  	& 75.8 	& 11.33 	& 22.85 	& 25.07 	& 2.22 \\ 
02080+3725  	& N834  	& \emph{02 11 01} 	& \emph{37 39 58} & 4593  	& 64.7 	& 10.95 	& 22.49 	& 24.69 	& 2.20 \\ 
02152+1418  	& N877  	& 02 17 58.62 	& 14 32 25.90 & 3913  	& 54.2 	& 10.98 	& 22.58 	& 24.74 	& 2.17 \\ 
02208+4744  	&   	& 02 24 08.00 	& 47 58 10.68 & 4679  	& 66.5 	& 11.09 	& 22.51 	& 24.84 	& 2.33 \\ 
\enddata
\label{ta:T1stub}

\begin{minipage}[t]{0.90\hsize}
{\sc Columns---}
(1): IRAS Name (from Point Source Calalog, excepting  F07258+3357 which is 
taken from the Faint Source Catlog).
(2):  Other name.
(3)-(4): Right Ascencion and Declination (from VLA observations, unless undetected in which case
IRAS positions are given in italics).  VLA positions are conservatively estimated to be accurate 
to 0.3$''$ and IRAS positions to 20$''$. 
(5): Heliocentric velocity (cz) from NED.
(6): Luminosity distance (see Section 2).
(7): FIR luminosity (see Section 3.3).
(8): 1.4~GHz luminosity calculated from the NVSS catalog \citep{NVSS}.
(9): FIR spectral luminosity (see Section 3,3).
(10): $q_{\nu}=\log(L_{\rm FIR}/L_{1.4})$.
The full version of this table is given in Table\,\ref{ta:T1full}.
\end{minipage}

\end{deluxetable}

\begin{deluxetable}{clcccc}
\tablenum{2}
\tablecolumns{6}
\tablewidth{0pt}
\tablecaption{Summary of Radio Observations}
\tablehead{
\colhead{Date}&
\colhead{Array}&
\colhead{Frequency}&
\colhead{Beamsize\tablenotemark{b}}&
\colhead{Sources}&
\colhead{Typical RMS}
\\
\colhead{}&
\colhead{}&
\colhead{[GHz]}&
\colhead{[mas]}&
\colhead{}&
\colhead{[\uJy~\per{beam}]\tablenotemark{d} }
}
\startdata
  98/06/23   &VLA BnA                     &4.9 & 900x400&$0^\circ<\delta\leq30^\circ$ &\phn75\\
  03/07/19   & VLA A                    &4.9 & 400 & $\delta>30^\circ$ and CC \tablenotemark{a}   &\phn75\\
  05/02/27   &Eb--Wb--Ar\tablenotemark{c} &4.9 & 1.8&$10^\circ<\delta\leq40^\circ$  &\phn 25\\
  05/05/06   &Eb--Wb--Jb1\tablenotemark{c}&4.9 & 18.1&$0^\circ<\delta\leq10^\circ$ and $\delta>40^\circ$  & 56\\
\enddata
\label{ta:obslist}
\tablenotetext{a}{CC=Sources known to be compact at 4.9~GHz from the
  first VLA epoch.}\tablenotetext{b}{For the VLA observations this is
  the typical FWHM of the restoring beam. For the VLBI observations
  this is the typical fringe spacing of the longest baseline.}
\tablenotetext{c}{Eb=Effelsberg, Wb=Westerbork, Ar=Arecibo and
  Jb1=Jodrell Bank (Lovell). Baselines lengths are Eb--Wb=266~km,
  Eb--Jb1=699~km and Eb--Ar=6901~km.}
 \tablenotetext{d}{r.m.s noise has units of \uJy~\per{beam} for VLA observations
 but for VLBI observations refers to noise on most sensitive baseline in  \uJy.
 }

\end{deluxetable}

\subsection{Archival Data Radio and IR data} \label{se:archival} In
addition to our own radio data our analysis made use of source
1.4\,GHz total flux densities taken from the \NVSS\ catalog
\citep{NVSS}.  Infra-red total flux densities were taken from the
\IRAS\ Point Source Catalog \citep{IRAS88}. In order to calculate
source infrared to radio (1.4\,GHz) ratios ($q_{1.4}$) as defined by
\citet{HELOU85} we calculated for each source an equivalent FIR flux
density from $S_{\rm
  FIR}=1.26\times10^{-14}(2.58f_{60}+f_{100})/3.75\times
10^{12}$~W\,m$^{-2}$\,\per{Hz} where $f_{60}$ and $f_{100}$ are the
cataloged 60 and 100\,\um\ \IRAS\ flux densities. For comparison with
COLA-S and other data sets for each source a FIR spectral luminosity
(see Table~\ref{ta:T1stub}, column 9) was calculated assuming
isotropic emission. Finally bolometric IR (8$\mu$m -1000$\mu$m)
luminosities in solar units were calculated (see
Table~\ref{ta:T1stub}, column 7) from the IRAS flux densities using
the relation given by \citet{SANDERS96}.

\subsection{Optical spectroscopy observations}
\label{optobs} 

Optical spectroscopy observations were taken using the 60 inch CfA
telescope at Mt Hopkins, Arizona between February and October 2001
using the FAST spectrograph. A slit width of 3\arcsec\ was used,
corresponding to 0.9\,kpc at the median sample source distance. A
grating of 300 lines/mm was used to disperse light between 4000 and
7000\AA\ and gave a spectral resolution of 6\AA. A single exposure of
between 900 and 1200 seconds was made for each
target. Spectrophotometric and smooth standards (often the same stars)
were observed during each night and used to flux-calibrate and correct
for atmospheric absorption features in the spectra.  We were able to
measure useful line ratios for 61\% of the sample.

Data reduction was carried out in the standard manner using IRAF. The
frames were de-biased and flat-fielded. Galaxy spectra were extracted
using an aperture centered on the position of peak flux across the
slit. If more than one peak was seen in the spatial direction a
spectrum was extracted at the position of each peak. Sky-subtraction
was then carried out and wavelength calibration done using an Ne-Ar
arc lamp. The observations were then corrected for atmospheric
extinction and flux calibrated using the spectrum of a
spectrophotometric standard. Finally atmospheric absorption features
in the target spectra were removed using the smooth spectrum standard
observed at a similar airmass to the target.

The line fluxes were measured by fitting (multiple) Gaussians to the
emission lines using NGAUSSFIT in IRAF. The continuum flux was
subtracted using a linear fit to small portions of the continuum
either side of the emission line. The H$\alpha$ ($\lambda$ 6562\AA)
and [NII] doublet ($\lambda\lambda$ 6583, 6548\AA) were often
overlapping, in these cases line profiles were de-blended by
simultaneously fitting a separate Gaussian for each line.  When
H$\alpha$ was present in absorption as well as emission both the
emission and absorption features were fitted simultaneously with
Gaussians.

\section{Results}

\begin{figure*}
  \centering
  \includegraphics[width=1.0\hsize]{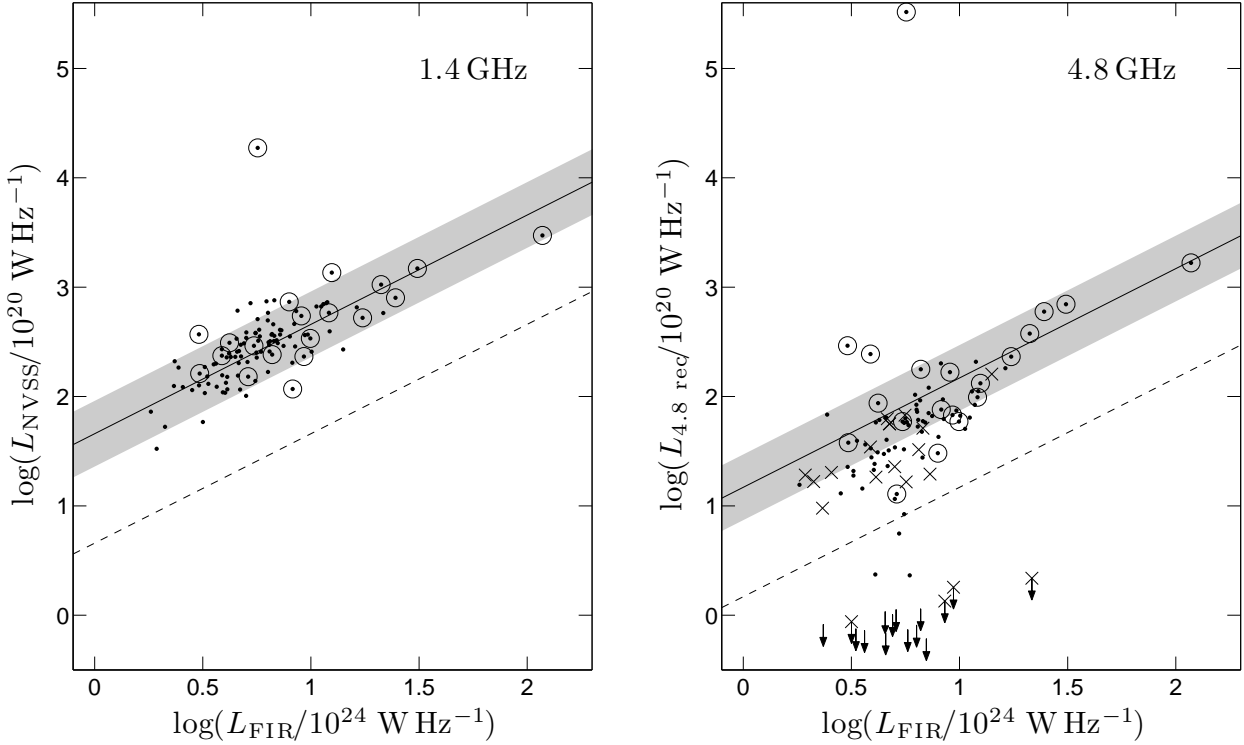}\hfil
  \caption{FIR to Radio correlation plots for the COLA North sample at
    1.4\,GHz and at 4.8\,GHz. In both panels sources detected by VLBI
    are shown circled.  Left: 1.4\,GHz radio spectral luminosity
    calculated using \NVSS\ fluxes versus FIR spectral luminosity
    $L_{\rm FIR}$ (see \S\ref{se:archival}). Most of the COLA North
    sources are tightly clustered around the median $q_{1.4}=2.34$
    estimated for a sample of 1809 galaxies by \citet{YUN01} which is
    shown by a solid line. Note on average that for the sample $L_{\rm
      FIR}= 1\times10^{25}$ WHz$^{-1} $ corresponds to $L_{\rm IR}=
    10^{11.25} \Lsun $.  Right: 4.8~GHz spectral luminosity calculated
    using the recovered 4.8\,GHz fluxes from our VLA maps (see
    \S\ref{se:VLA}) versus $L_{\rm FIR}$.  Arrows indicate $5\sigma$
    upper limits for the 15 sources not detected in these
    observations. Crosses indicate the 20 sources having no VLBI
    data. The median $q_{4.8}=2.83$ measured by \citet{WUNDERLICH87}
    from a sample of 99 normal spirals is shown by a solid line. In
    both panels, the gray area defines the region of $\pm0.3$\,dex
    around the median and a dashed line is drawn at $-1$\,dex.}
  \label{fi:radio_fir}
\end{figure*}

\begin{figure*}[t!]
  \centering
  \includegraphics[width=0.8\hsize]{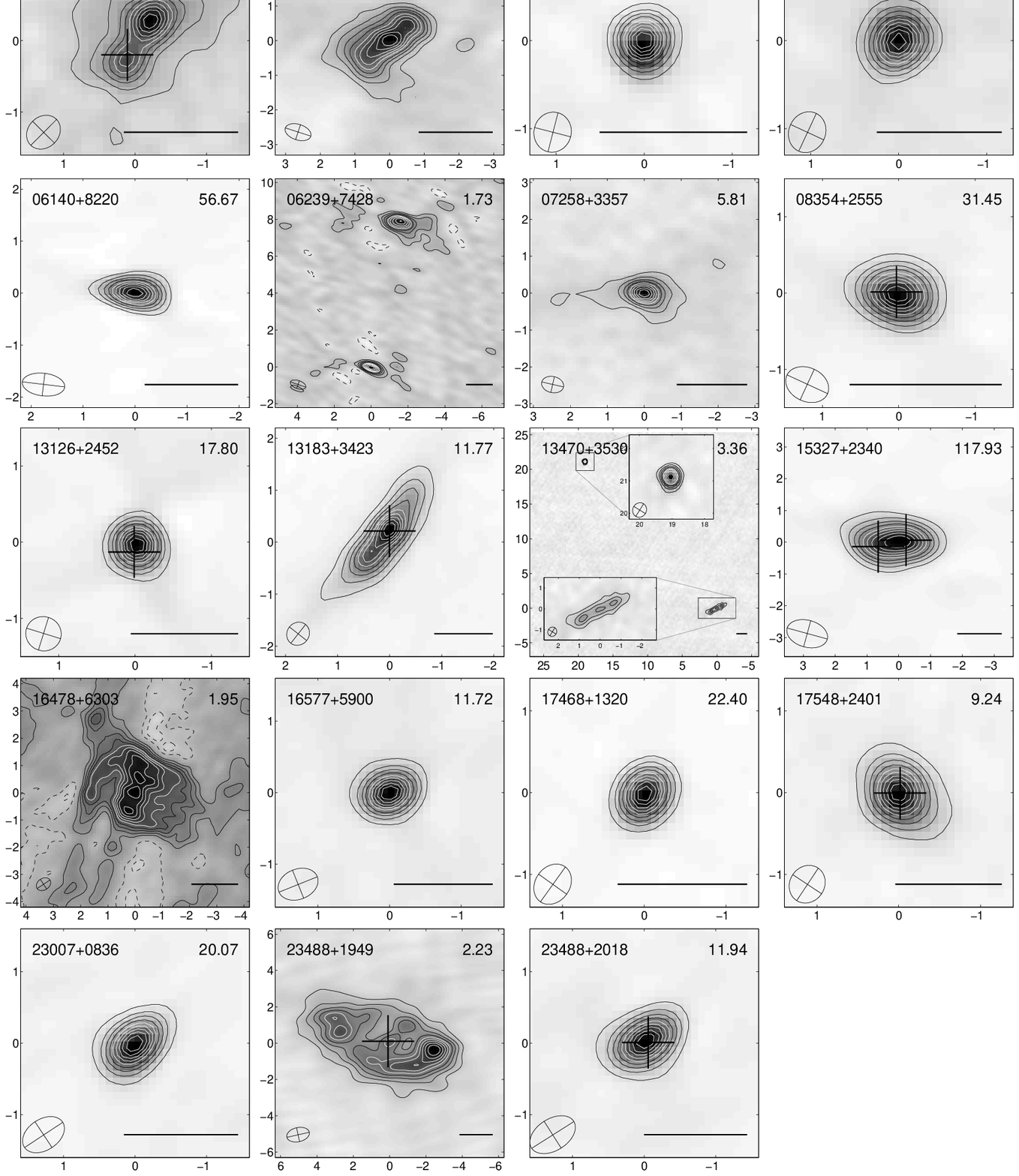}
  \caption{\scriptsize VLA 4.8\,GHz naturally weighted maps for 19 of
    the 20 VLBI detected sources (the well known core-dominated source
    031643+4119 (3C84) is omitted).  VLA images of the sources
    undetected by VLBI are given in Figure~\ref{fi:nondet1}.  Shown
    are the highest resolution images of each source which in all
    cases come from A-array observations except for 15327+2340
    (Arp220) and 23488+1949 which come from BnA observations. In
    05336+5407 two features are detected 30\arcsec\ apart, only the
    brighter more compact NW feature which contains the VLBI source is
    shown.  In all of the figures the IRAS source name is indicated in
    the top left-corner. Peak brightness in mJy\,\per{beam} is
    indicated in the top right corner. Black contours are drawn at 10,
    20, 30 and 40\% of the peak brightness. White contours are drawn
    at 50, 60, 70, 80 and 90\% of the peak brightness. Negative
    contours are drawn with dashed lines at $-$10, $-$20, $-$30 and
    $-$40\% of the peak brightness. A cross indicates the position of
    the VLBI core detections for those source having observations
    using the Eb--Ar baseline.  Axes tick-marks are in arcseconds. The
    horizontal line on the bottom-right corner indicates 500\,pc at
    the distance of the source.}
  \label{fi:vla_maps_detections}
\end{figure*}

\subsection{Sample overall Radio/FIR ratios}\label{se:sample-q}
In a FIR-selected sample of galaxies such as COLA, it is expected that
in most sources star-formation will be the dominant mechanism powering
the FIR emission. For such sources radio and FIR luminosities and flux
densities are expected to be tightly correlated
\citep[see][]{CONDON_REVIEW92}. The total 1.4\,GHz and FIR
luminosities for our sample are plotted in
Figure\,\ref{fi:radio_fir}-Left. Except for one outlier (3C84) which
is unusually radio luminous, sources tightly follow the expected
correlation. We calculated for each source the logarithmic ratio of
flux densities $q_{1.4}=\log(S_{\rm FIR}/S_{1.4})$ as defined by
\citet{HELOU85}. Excluding 3C84, the mean value of $q_{1.4}$ is $2.34
\pm 0.01$ (and the median $2.31 \pm 0.01$) which is in good agreement
with the mean $q_{1.4} = 2.34 \pm 0.01$ measured by \citet{YUN01} from
a sample of 1809 galaxies (solid line in Figure~\ref{fi:radio_fir}).\
In contrast the median value for COLA-S \citep{CORBETT02} is $q_{1.4}
= 2.43\pm0.03$ which is significantly (4$\sigma$) larger than the
COLA-N median value.
 
To try to resolve the above discrepancy we compared \NVSS\ and COLA
fluxes for the 56 COLA~South sources also contained in the \NVSS\
catalog (i.e.\ those with $\delta\geq-40^{\circ}$). We found \NVSS\
flux densities for these sources which were 12\% larger than those
measured by \citet{CORBETT02}, resulting in $q_{1.4} = 2.36\pm0.05$
for ATCA fluxes and $2.31\pm0.04$ for NVSS fluxes. It has been noted
by \citet{PRANDONI00}, \citet{NOR06} and \citet{KELLERMANN08} that for
sources with flux densities $\lesssim15$\,mJy \NVSS\ flux densities
tend to be $\about$ 10\% larger than ATCA measurements. Although COLA
total flux densities are about a factor of two above this flux limit
the effect is neither understood nor well-defined and so this scale
difference may contribute to the apparent difference of median
$q_{1.4}$. We conclude that the difference in $q_{1.4}$ between COLA-N
and COLA-S is likely caused by (a) a $\sim$10\% calibration difference
for weak sources between NVSS and ATCA flux densities, combined with
(b) a small number of sources at $\delta<-40^{\circ}$ which have
anomalously large values of $q_{1.4}$, perhaps simply because of small
number statistics.

It remains unclear whether the likely $\sim$10\% calibration
difference between Northern and Southern hemispheres noted above for
weak sources is a property of NVSS specifically or all VLA
observations. Furthermore the 4.8\,GHz measurements of galaxies made
with the Bonn 100m single dish give $q_{4.8}=2.71\pm0.03$
\citep{WUNDERLICH87}, assuming a spectral index of 0.75 this then
gives an estimated $q_{1.4}=2.31\pm0.03$ close to the VLA value,
supporting the idea that Northern hemisphere $q_{1.4}$ values are
consistent with each other. Given the importance of sorting out the
origin of radio flux scale differences between hemispheres it seems
that dedicated simultaneous ATCA/VLA observations of sources with
different flux densities should be made.

\begin{figure*}[t!]
  \figurenum{3.1--3.7} \centering
  \includegraphics[width=0.7\hsize]{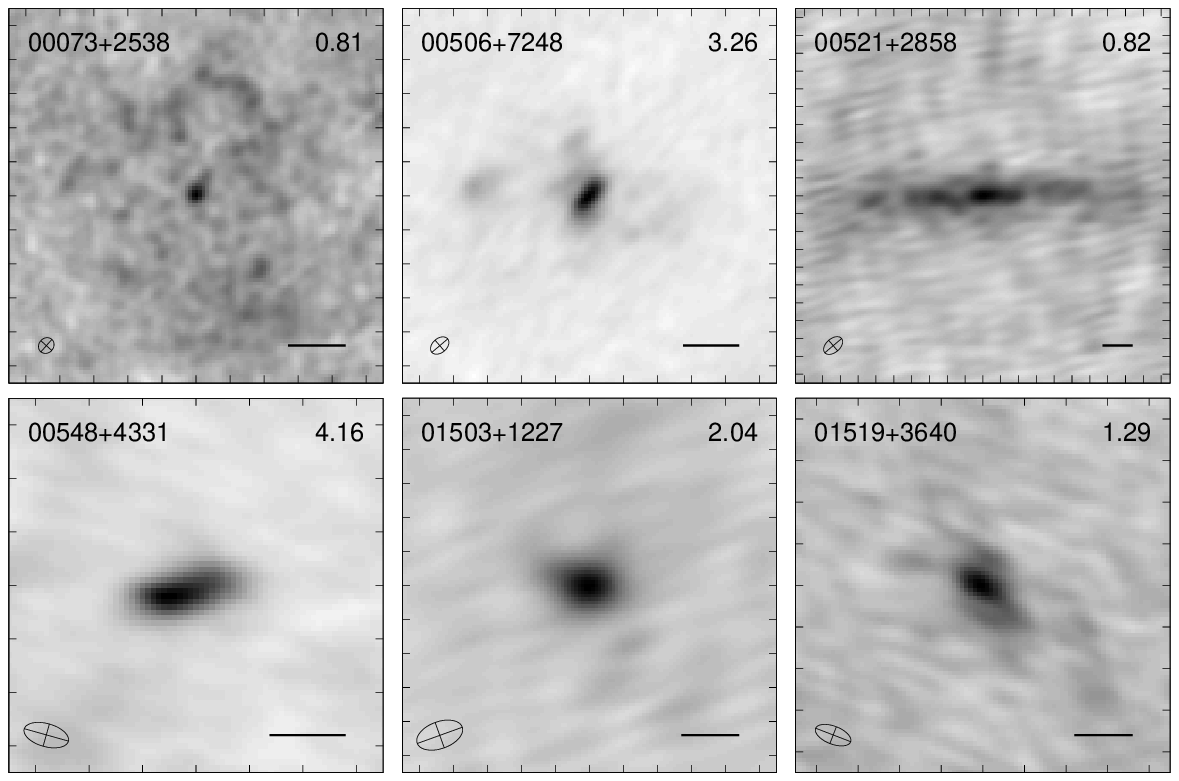}
  \caption{VLA 4.8\,GHz naturally weighted maps for the VLA detected
    sources of the COLA North sample without VLBI detections. Maps for
    the VLBI detections are given in
    Figure~\ref{fi:vla_maps_detections}. Column~2 in Table\,3 gives
    the VLA array used, for sources observed in two arrays the higher
    resolution image is shown.  IRAS names are indicated on the
    top-left corner of every panel. Peak brightness in mJy~\per{beam}
    is indicated on the top-right corner.  Tick marks are separated by
    1 arcsecond with the restoring beam shown at the bottom-left
    corner.  The horizontal line at the bottom-right corner is 500~pc
    at the distance of the source. Only the first six images are shown
    here the remaining images can be found in the online version
    Figures\,3.2 to 3.7. }
  \label{fi:nondet1}
\end{figure*}


\begin{deluxetable}{ c c c c c c c c c c }
\centering
\tablecolumns{10}
\tablenum{3}
\tabletypesize{\scriptsize}
\tablewidth{0pt}
\tablecaption{\sc Radio and optical observations and results}
\tablehead{
\colhead{IRAS}&
\colhead{VLA }&
\colhead{VLA }&
\colhead{$S_{\rm 4.8~total}$}&
\colhead{$q_{4.8}$ } &
\colhead{$S_{\rm 4.8~comp}$}&
\colhead{$\theta_{M}\x\theta_{m}$}&
\colhead{$\log T_b$}&
\colhead{VLBI} &
\colhead{Optical}
\\
\colhead{Name}&
\colhead{Array}&
\colhead{Morph}&
\colhead{(mJy)}&
\colhead{}&
\colhead{(mJy)}&
\colhead{[\arcsec\x\arcsec]}&
\colhead{[K]}&
\colhead{Epoch} &
\colhead{Class}
\\
\colhead{(1)}&
\colhead{(2)}&
\colhead{(3)}&
\colhead{(4)}&
\colhead{(5)}&
\colhead{(6)}&
\colhead{(7)}&
\colhead{(8)}&
\colhead{(9)}&
\colhead{(10)}
\\
}
\startdata
00005+2140 	& BnA--A 	& CE 	& 5.66 	& 2.97 	& 0.00 	& -- 	& -- 	& 1*  	& S  \\ 
00073+2538 	& BnA--A 	& C 	& 18.40 	& 2.85 	& 1.23 	& 0.45\x0.22 	& 2.63 	& 1   	& C  \\ 
00506+7248 	& A 	& CE 	& 31.60 	& 2.95 	& 13.60 	& 1.27\x0.66 	& 2.75 	& 2   	& --  \\ 
00521+2858 	& BnA 	& E 	& 6.07 	& 3.13 	& 3.40 	& 3.15\x0.91 	& 1.61 	& 1   	& --  \\ 
00548+4331 	& A 	& CE 	& 13.21 	& 2.95 	& 13.11 	& 1.51\x0.41 	& 2.87 	& 2   	& --  \\ 
00555+7614 	& A 	& U 	& -- 	& -- 	& -- 	& -- 	& -- 	& 2   	& --  \\ 
01503+1227 	& BnA 	& CE 	& 6.23 	& 3.18 	& 5.25 	& 1.58\x1.03 	& 2.05 	& 1   	& C  \\ 
01519+3640 	& A 	& CE 	& 4.38 	& 3.23 	& 3.92 	& 1.13\x0.74 	& 2.21 	& 1-2 	& C  \\ 
01555+0250 	& BnA 	& CE 	& 8.23 	& 2.92 	& 1.85 	& 2.04\x1.06 	& 1.47 	& --  	& L  \\ 
01556+2507 	& BnA 	& CE 	& 11.16 	& 2.95 	& 7.37 	& 1.32\x1.02 	& 2.28 	& 1   	& H  \\ 
01579+5015 	& A 	& CE 	& 2.49 	& 3.39 	& 1.82 	& 0.59\x0.38 	& 2.45 	& 2   	& L  \\ 
02071+3857 	& A 	& E 	& 12.20 	& 3.15 	& 0.93 	& 0.85\x0.56 	& 1.83 	& 2   	& --  \\ 
02080+3725 	& A 	& U 	& -- 	& -- 	& -- 	& -- 	& -- 	& 1-2 	& H  \\ 
02152+1418 	& BnA 	& CE 	& 2.39 	& 3.82 	& 1.34 	& 1.49\x0.94 	& 1.52 	& 1-2 	& --  \\ 
02208+4744 	& A 	& CE 	& 10.92 	& 3.08 	& 10.61 	& 1.78\x1.02 	& 2.31 	& 2   	& --  \\ 
02253+1922 	& BnA 	& U 	& -- 	& -- 	& -- 	& -- 	& -- 	& 1-2 	& H  \\ 
\enddata
\label{ta:T3stub}

\begin{minipage}[t]{0.90\hsize}
{\sc Columns---}
(1): IRAS Name
(2): VLA array(s) used
(3): Morphology class of VLA 4.8GHz image. C-Compact,  E-Extended,  CE-Compact plus Extended, CER-Compact plus extended with ring, U-Undetected.
(4): Recovered 4.8GHz flux density in VLA image.
(5). Log of ratio of 4.8GHz recovered VLA flux density  to FIR flux density.
(6): Total flux density of gaussian fitted to brightest feature in VLA image (after removing VLBI component, hence 
 zero indicates that brightest feature was dominated by the VLBI component).  
  (7): Fitted gaussian major and minor axes to peak VLA feature (FWHM arcsec), Note 
  for  IRAS15327+2340 (Arp~220)  these are for the western nucleus as taken from the literature (see Section  \ref{se:jmfit}).
(7): Fitted gaussian major and minor axes to peak VLA feature (FWHM arcsec).
(8): Peak  4.8GHz brightness temperature of gaussian component  (see Section  \ref{se:jmfit}). Note if formally unresolved along one axis 
a 50mas FWHM was assumed along that axis).
(9): VLBI Epoch(s) observed (asterisk indicates VLBI detection, see Table~\ref{ta:VLBI_summary} for details).
(10): Optical Class  (see Section \ref{se:optclass})  with {\tt H}=HII, {\tt S}=Seyfert, {\tt
    L}=LINER and {\tt C}=Composite.
\end{minipage}

\end{deluxetable}

\subsection{VLA 4.8\,GHz Maps}\label{se:VLA_maps}
The results of our VLA observations are summarized in Table\,3. We
successfully detected and mapped 95 out of the 110 observed
sources. The 15 non-detections are probably large galaxies with a
surface brightness too low to be detected with the VLA configuration
used.

The images for the VLA detections are shown in
Figure~\ref{fi:vla_maps_detections} and Figures~\ref{fi:nondet1} with
the former showing sources which are also VLBI detections and the
latter showing those that are VLBI non-detections. The images show a
wide variety of morphologies. Following \citet{NEFF92} we have
classified the sources into four different morphological classes: Core
(C), Core and Extensions (CE), Extended with no core (E) and
Undetected (U) each class containing 28, 62, 5 and 15 sources
respectively. There are two sources with clear ring structures, namely
IRAS05091+0508 (NGC\,1819) and 23488+1949 (NGC\,7771), there are an
additional four showing elongated linear structures typical of edge-on
disks.

Total recovered fluxes at 4.8\,GHz ($S_{\rm 4.8~total}$) were obtained
by integrating the flux over a region enclosed by the lowest reliable
contour of radio emission. The median $q_{4.8}=\log(L_{\rm FIR}/L_{\rm
  4.8~tot})$ obtained for the sample is $3.03\pm0.05$ which is
significantly higher than the $2.71\pm0.03$ measured by
\citet{WUNDERLICH87} using single dish observations of a sample of 99
normal spirals. We conclude that we are on average resolving
out \about50\% of the 4.8\,GHz flux density. We note however that for
those sources with compact central features the fraction of missing
flux density is $\lesssim20\%$.

\begin{deluxetable}{ccrrrccc}
\tabletypesize{\scriptsize}
\tablenum{4}
  \tablewidth{0pt}
  \tablecolumns{8}
  \tablecaption{VLBI Results}
    \tablehead{
    \colhead{COLA}&
    \colhead{Other}&
    \colhead{Eb--Wb}&
    \colhead{Eb--Jb1}&
    \colhead{Eb--Ar}&
    \colhead{Angular Size}&
    \colhead{Linear Size}&
    \colhead{Optical}
\\
    \colhead{name}&
    \colhead{name}&
    \colhead{[mJy]}&
    \colhead{[mJy]}&
    \colhead{[mJy]}&
    \colhead{[mas]}&
    \colhead{[pc]}&
    \colhead{Class}
\\
    \colhead{(1)}&
    \colhead{(2)}&
    \colhead{(3)}&
    \colhead{(4)}&
    \colhead{(5)}&
    \colhead{(6)}&
    \colhead{(7)}&
    \colhead{(8)}
}
\startdata
 00005+2140  &Mrk 334&1.67 (0.15)     &\NO	   &1.02 (0.03) &  0.68 & 0.29                               & {\tt S}\\
  02568+3637     &       &1.61 (0.14)     &1.92 (0.16)  &\NO 	&       &                                    & {\tt S}\\
  03164+4119      &   3C84&                &	           &15500(0.02) &       &                                    & {\tt L}\\
  04435+1822 &       &3.32 (0.14)     &\NO	   &3.10 (0.03) &  0.25 & 0.07                               & {\tt H}\\
  05054+1718&       &1.64 (0.18)     &\NO	   &$<0.20$	& $>$1.41\phantom{$>$} &$>$0.50\phantom{$>$} & {\tt S}\\
  06140+8220  &       &4.12 (0.32)     &\NO	   &\NO 	&       &                                    & {\tt -} \\
  06239+7428   &       &\NO	     &1.51 (0.19)  &\NO 	&       &                                    & {\tt -} \\
  F07258+3357  &  N2388&1.38 (0.18)     &1.26 (0.14)  &$<0.20$	& $>$1.35\phantom{$>$} &$>$0.35\phantom{$>$} & {\tt H}\\
  08354+2555 &Arp 243&2.03 (0.10)     &\NO	   &0.27 (0.02) &  1.38 & 0.49                               & {\tt  -}\\
  13126+2452   &       &1.38 (0.17)     &\NO	   &0.88 (0.04) &  0.65 & 0.16                               & {\tt - }\\
  13183+3423   &Arp 193&2.17 (0.17)     &3.85 (0.15)  &0.70 (0.04) &  1.03 & 0.46                               & {\tt C}\\
  13470+3530  &       &1.79 (0.11)     &5.64 (0.16)  &$<0.18$	&  $>$1.47\phantom{$>$} &$>$0.48\phantom{$>$}& {\tt -}\\
  15327+2340   &Arp 220&4.37 (0.17)     &\NO	   &2.68 (0.04) &  0.40 & 0.14                               & {\tt L}\\
            &       &                &\NO	   &1.02 (0.04) &       &                                                   \\
  16478+6303   &  N6247&2.60 (0.12)     &$<0.82$      &\NO 	&       &                                    & {\tt -}\\
  16577+5900   &Arp 293&6.11 (0.27)     &5.36 (0.17)  &\NO 	&       &                                    & {\tt C }\\
  17468+1320   &       &8.33 (0.21)     &5.31 (0.13)  &$<0.25$	&  $>$1.82\phantom{$>$} &$>$0.57\phantom{$>$}& {\tt -}\\
  17548+2401 &       &4.45 (0.20)     &\NO	   &0.39 (0.04) &  1.52 & 0.58                               & {\tt L }\\
  23007+0836  &  N7469&5.12 (0.22)     &6.32 (0.15)  &\NO 	&       &                                    & {\tt S }\\
  23488+1949  &  N7771&$<1.14$         &\NO	   &0.33 (0.04) &  $<$1.08\phantom{$<$} &$<$0.28\phantom{$<$}& {\tt L }\\
  23488+2018   &Mrk 331&5.02 (0.21)     &4.92 (0.13)  &0.30 (0.04) &  1.63 & 0.57                               & {\tt C }\\
\enddata

\tablecomments{{\sc Columns} --- (1): IRAS Name (from Point Source Calalog, excepting  F07258+3357 which is 
taken from the Faint Source Catlog). (2): Other name.
  (3)-(5): VLBI flux and 1$\sigma$ uncertainty  (in brackets) obtained from the peak
  in the Delay--Rate map. Upper limits for the non-detections are
  indicated at $5\sigma$.  A dash is shown for sources not observed at
  that baseline. For IRAS15327+2340 (Arp~220) on the longer baselines the two rows give 
  results for the West and East nuclei  respectively.  (6): FWHM of a circular Gaussian source fitting the
  flux ratio between the Eb--Wb and Eb--Ar baselines (see
  \S\ref{se:vlbi_det}).  Sources with flux upper limits on the Eb--Ar
  have consequently size lower limits.  (7): Corresponding fitted
  Gaussian FWHM linear sizes. (8): Optical Class (see Section \ref{se:optclass}) with {\tt H}=HII, {\tt S}=Seyfert,  {\tt L}=LINER and {\tt C}=Composite; a dash indicates 
no optical line data.}
\label{ta:VLBI_summary}
\end{deluxetable}


\subsection{VLBI Detections}\label{se:vlbi_det}

\begin{figure*}[t!]
  \figurenum{4} \centering
  \includegraphics[width=0.7\hsize]{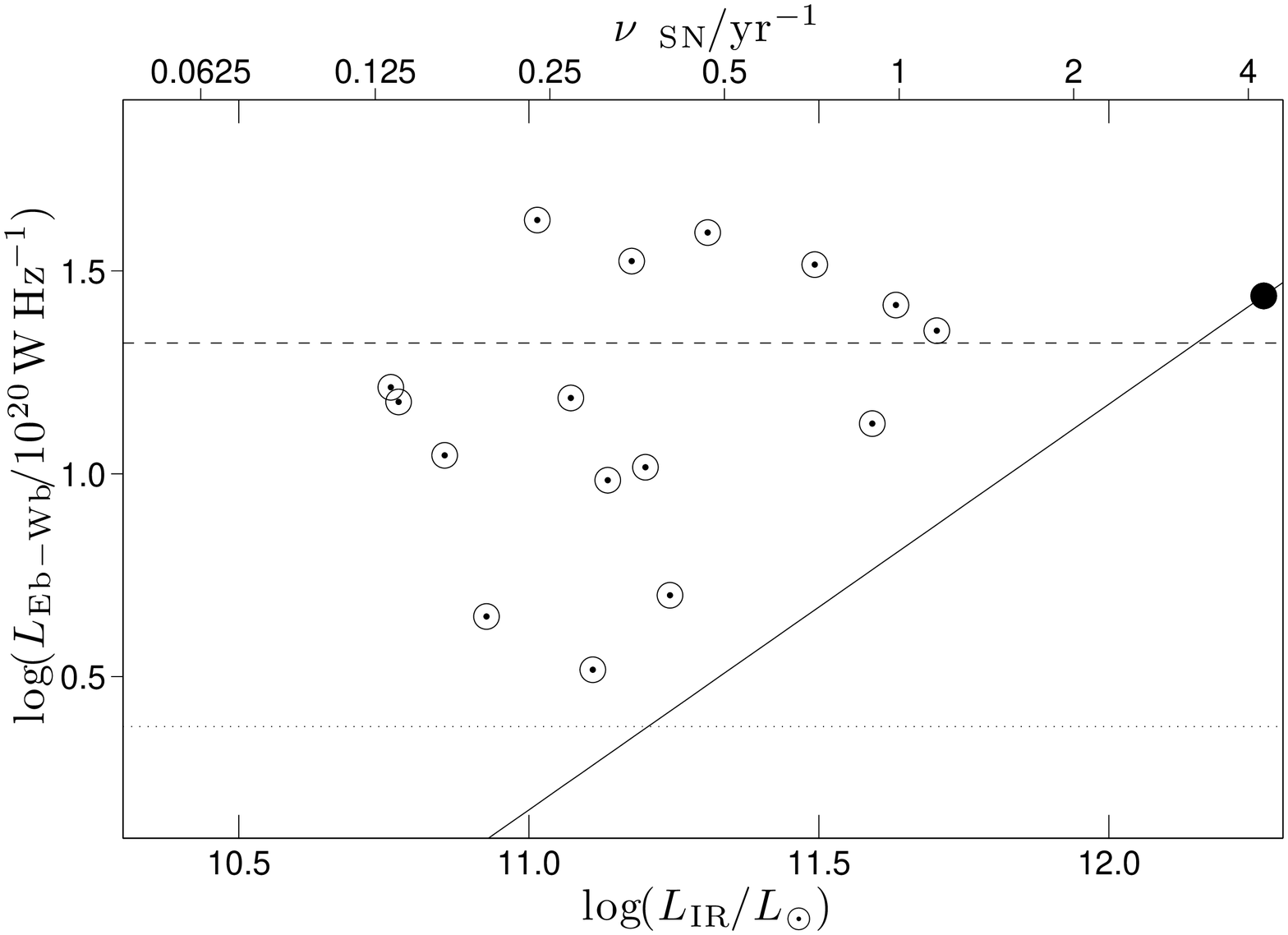}\hfil
  \caption{Radio spectral luminosity at 4.8\,GHz on the Eb--Wb
    baseline for VLBI detected sources (excluding 3C84) versus IR
    luminosity. The source Arp\,220 is shown as a filled circle. The
    scale on the upper horizontal axis indicates the core-collapse
    supernova rate $\nu_{\rm SN}$ estimated from the IR luminosity
    using the formula in \citet{VANBUREN94} (after applying a small
    FIR to IR luminosity correction).  The luminosity at maximum light
    measured for the powerful Type~IIn SN1986J is indicated with a
    horizontal dashed line at 2.1\x\E{21}\,W\,\per{Hz}.  The diagonal
    line represents the spectral luminosity detected on the Eb--Wb
    baseline for Arp\,220 scaled by IR luminosity. The lower
    horizontal dotted line indicates our typical 6$\sigma$=780\,\uJy\
    detection threshold on this baseline at the median distance of the
    COLA sample. }
  \label{fi:q_hist}
\end{figure*}

We detect, on at least one VLBI baseline, 20 out of 90 sources
observed. The measured flux densities on each baseline are given in
Table~\ref{ta:VLBI_summary}. On the shortest Eb--Wb baseline (see
Figure~\ref{fi:q_hist}) the signal to noise ratio for most detections
is generally significantly larger than the 6$\sigma$ detection
threshold (see Figure~\ref{fi:q_hist}) while none detections are
bounded to $<$4$\sigma$.  It is notable that we obtain VLBI detections
for 5 of the 6 sources with IR luminosities above \E{11.4}\,\Lsun\ .
A histogram showing the VLBI detections and non-detections versus IR
luminosity is presented in Figure~\ref{fi:Tb_hist}-top which shows
that the probability of detecting a VLBI core increases with IR
luminosity.

Table~\ref{ta:VLBI_summary} shows that there are 9 sources with
simultaneous measurements on the Eb--Wb and Eb--Jb1 baselines. Of
these, 7 have similar fluxes (within the noise) on these two
baselines, the latter of which is \about2.6 times longer than the
first, showing that the VLBI component is unresolved.

Of the 14 sources with simultaneous Eb--Wb and Eb--Ar observations 9
are detected on the Eb--Ar baseline, but with significantly reduced
flux density, indicating that the VLBI sources are resolved on the
longest baseline. We estimate an angular size (see
Table~\ref{ta:VLBI_summary}) by approximating the flux ratio between
both baselines to that expected from a circular Gaussian
\citep[see][]{PEARSON99}.

\subsection{FIR/radio ratios of the VLBI Detections}\label{se:vlbi-q}

Figure~\ref{fi:radio_fir}-left shows that the total 1.4\,GHz VLA flux
densities of our VLBI detected sources follow the FIR to radio
correlation. There are 15/20 detections within $\pm0.3$~dex of the
empirical line defined by \citet{YUN01} (i.e. within the grey region
shown).
Of the remainder two are just slightly above and two slightly below
the $\pm$0.3\,dex line. Only one source, 3C84, is significantly above
the correlation. At 4.8\,GHz a similar trend is observed with 14/20
detections located within $\pm0.3$\,dex of the empirical line defined
by \citet{WUNDERLICH87}. Consistent with the above results we find the
median $q_{1.4}$ for the VLBI detections and non-detections to be
indistinguishable (i.e. $2.30\pm0.08$ and $2.31\pm0.02$
respectively). Likewise if we omit 3C84 the mean of the VLBI detection
is $2.33\pm0.08$ and of the non-detections is $2.32\pm0.02$. We can
conclude that the data show that excepting 3C84 the total radio flux
density is dominated by star-formation regardless of whether or not a
VLBI core is present.
 
The above results appear to differ from those in COLA-S where a
difference in median $q_{1.4}$ between VLBI detections and
non-detections was claimed. From Table-3\ of \citet{CORBETT02}\ the
median $q_{1.4}$ value and its estimated one sigma error for the nine
definite COLA-S VLBI detections was $2.07\pm0.17$ while for the VLBI
non-detections it was $2.44\pm0.03$. The errors on theses estimates
are calculated from the internal dispersion about the median $q_{1.4}$
for the two sub-samples as given in \citet{CORBETT02}. We ascribe this
apparent two sigma difference between the two sub-samples in the South
as being due to the lower VLBI sensitivity in the South combined with
the overall form of the $q_{1.4}$ distribution. This distribution has
a tail of radio core-dominated, low $q_{1.4}$ sources (such as 3C84 in
the North and NGC5793 in the South) which are easily VLBI detected
plus weaker VLBI cores in galaxies whose radio emission is
star-formation dominated. The lower sensitivity VLBI observations in
the South contain a larger fraction of the former class of source
which reduces the estimated $q_{1.4}$ of VLBI detection compared to
non-detections. Specifically in the South 4 out of 9 VLBI detected
sources have $q_{1.4}\lesssim2$ while the fraction is only 3 out of 20
in the North. Because low $q_{1.4}$ sources comprise almost half of
the VLBI detected sample in the South they can bias the median
$q_{1.4}$ while in contrast in the North such outlier sources have
negligible effect.

\subsection{Relation Between VLA and VLBI Sources}\label{se:jmfit}
Visual inspection of the VLA radio maps indicates a bias for VLBI
detections to be in sources with compact structure and high peak
brightness (see Figure \ref{fi:vla_maps_detections}). This is
confirmed in terms of VLBI detection rate versus morphology class (see
\S\ref{se:VLA_maps}) .  Amongst the VLBI observed sources there are
13/27 detections of class C, 6/48 of class CE and 1/5 of class E. We
do not detect any class U source. This result was confirmed by
plotting VLA visibility amplitude versus $uv$ distance. The majority
of VLBI detections were indeed found in sources with compact, but not
unresolved, arcsec-scale structures. The size of these structures was
estimated by subtracting an unresolved source equivalent to the VLBI
detection from the brightest component in each VLA map and then fitting
a Gaussian to the difference.

In two cases (00005+2140 and 06239+7428) the VLBI core subtraction
removed all the flux density implying that the brightest VLA
component was caused completely by the VLBI source itself. We consider
these two sources together with 03164+4119 (3C84) to be \emph{VLBI
  core-dominated sources} and omit them from the subsequent
analysis. For the rest of the sources the task JMFIT was used to fit a
Gaussian within a tight box containing the remaining feature. For the
vast majority of sources in classes C and CE with a clear compact
component there was no ambiguity in the Gaussian fitting, i.e., the
results of the fits were the same regardless of the size of the
fitting box. For sources in class E the size estimates were found to
vary by up to a factor of \about2 depending on the chosen box. However
in all such cases the fitted sizes were large ($\gtrsim$2\arcsec).
The resulting component flux densities and major and minor axes
$\theta_M$ and $\theta_m$ are listed in columns 6 and 7 of Table\,3.

\begin{figure}
  \figurenum{5} \centering
  \includegraphics[width=0.9\hsize]{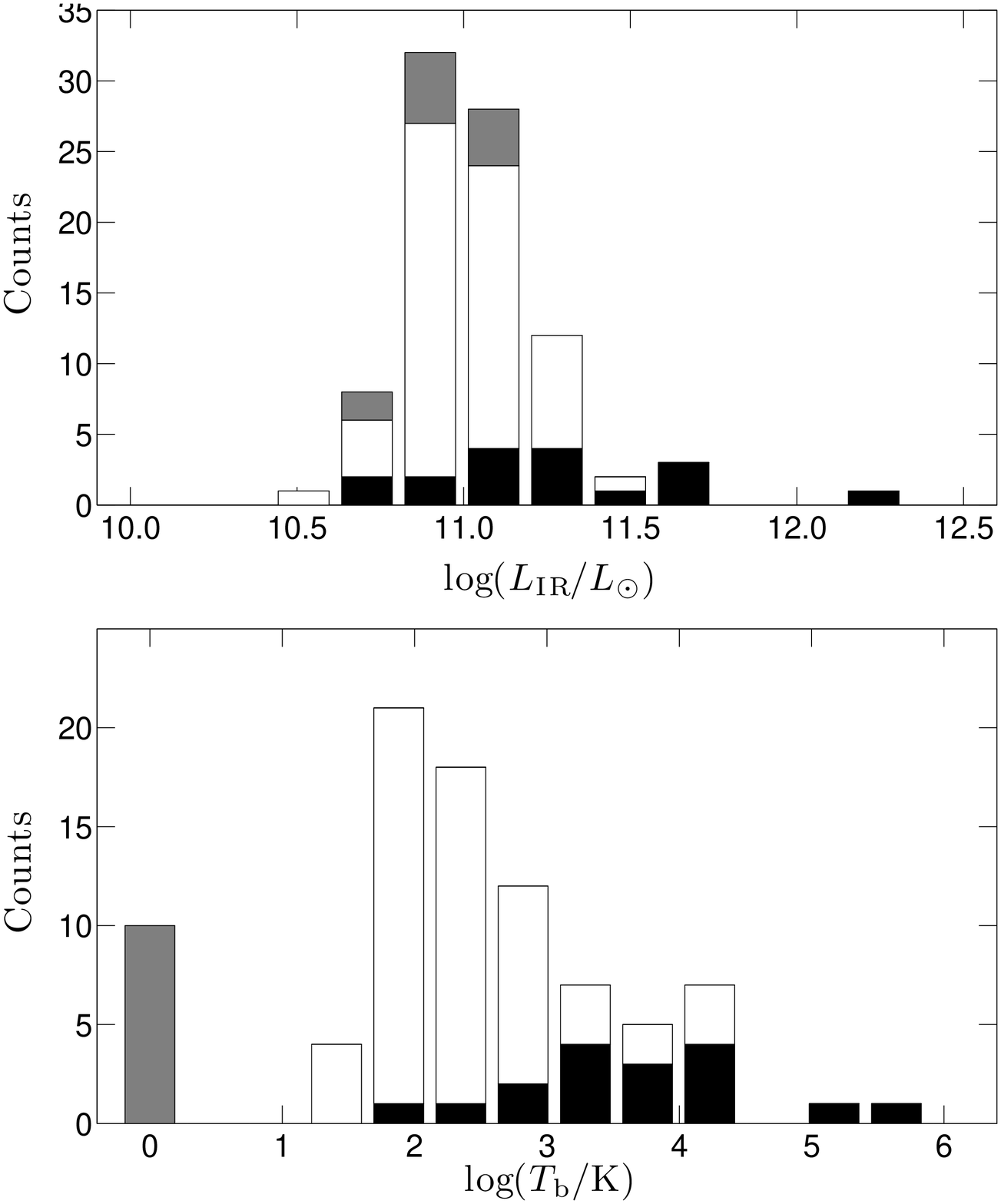}\hfil
  \caption{Histograms of IR luminosity (top panel) and peak brightness
    temperature of VLA scale radio component (bottom panel) for VLBI
    observed sources. In both panels VLBI detections are shown
    black. VLBI non-detections for sources respectively detected and
    undetected by the VLA are shown white and grey respectively. The
    brightness temperature in the bottom panel is calculated after
    subtraction of the VLBI component from the VLA map (see
    \S\ref{se:VLA_maps}).  Note for ease of comparison the three VLBI
    core-dominated detections (i.e 00005+2140, 03164+4119, 06239+7428) 
    are not plotted in either panel.  A much stronger correlation
    between VLBI detection and high brightness temperature VLA scale
    emission is found than with IR luminosity (see Section
    \ref{se:jmfit} and also Fig \ref{fi:todd} ) }
  \label{fi:Tb_hist}
\end{figure}

\begin{figure*}[t!] 
  \figurenum{6} \centering
  \includegraphics[width=0.99\hsize]{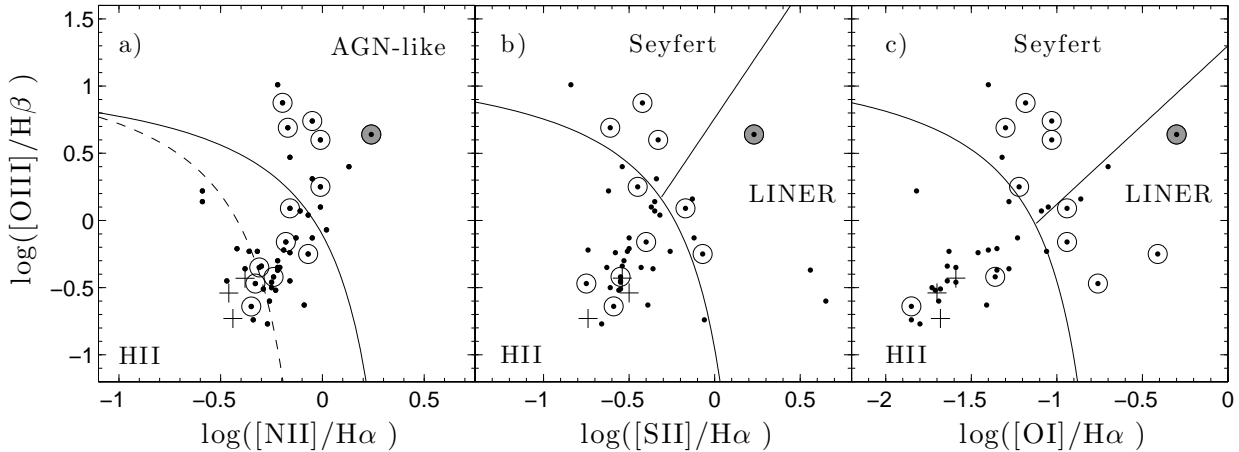}\hfil
  \caption{Optical-line diagnostic diagrams (``BPT'' diagrams) for the
    COLA-N sources that are both VLBI observed and have available
    optical line data.  Dots indicate VLA detected sources and crosses
    those undetected by the VLA.  VLBI detected sources are shown
    circled.  {\bf a) }[NII]/H$\alpha$ v/s [OIII]/H$\beta$ diagram.
    The solid and dashed curves are respectively the theoretical
    \cite{KEWLEY01} extreme starburst line and the \cite{KAUF03}
    empirical line for star-forming galaxies. Sources that lie between
    the two curves are classified as composite Starburst/AGN.  {\bf b)
    }[SII]/H$\alpha$ v/s [OIII]/H$\beta$ diagram {\bf c)
    }[OI]/H$\alpha$ v/s [OIII]/H$\beta$ diagram. In the latter two
    panels the solid curved line is the extreme starburst line from
    \cite{KEWLEY01} and the straight diagonal line separates Seyferts
    from LINERS \citep{KEWLEY06}. Note that for some sources not all
    line ratios are plotted because some lines were not detected.  The
    shaded VLBI detected source is 15327+2340 (Arp220).}
  \label{fi:BPTs}
\end{figure*}

Peak brightness temperatures $T_b$ at 4.8\,GHz for the compact VLA
structures were calculated using the relation $T_b\simeq35\x (S_{\rm
  4.8\,peak})(\theta_M\x\theta_m)^{-1}$~K$^{\circ}$ (Column\,8 in
Table\,3). In a few sources the fitted deconvolved minor axis was
zero, in these cases its size was set to 50\,mas which was equal to
the smallest size found in other sources.

Uniquely amongst sources with compact structure we did not collect any
VLA A-array data for the well known ULIRG source 15327+2340 (Arp220).
In order to treat this source in a similar way to the rest we give in
Table\,3 radio component properties from the literature for the
brighter western nucleus (sizes from \citet{CONDON91} with flux
densities scaled to 5\,GHz using the spectral index of
\citet{ROVILOS03}).

As can be seen from Figure~\ref{fi:Tb_hist} (bottom panel) VLBI
detections seem to be strongly correlated with those sources whose VLA
fitted components (after removing any VLBI contribution) have the
highest $T_b$. VLBI detectability is less strongly correlated with
large $L_{\rm IR}$ (top panel). Two sided Kolmogorov-Smirnov tests
comparing the distribution of VLBI detected and non-detected sources
in $L_{\rm IR}$ and $T_b$ in each case rejects the hypothesis that
VLBI/non-VLBI sources are drawn from the same population at
probability levels of $P=0.82$\ and $P=0.98$\ respectively. This
result demonstrates that the stronger correlation for VLBI
detectability is with $T_b$ rather than with $L_{\rm IR}$.

\subsection{Optical Spectroscopy Classifications}\label{se:optclass}

To classify sources based on their line ratios from our optical
spectroscopy observations (see \S\ref{optobs} ) we used the method of
\cite{KEWLEY06}. This scheme first classifies sources into HII
galaxies, AGN/star-formation composites and pure AGN based on a
mixture of theoretical \citep[\emph{maximal starburst,}][]{KEWLEY01}
and empirical line-ratio boundaries \citep{KAUF03} delineating pure
star-forming galaxies. These boundaries are shown respectively as
solid and dashed curves delimiting the bottom left corners of the
diagnostic line ratio diagrams in Figure\,\ref{fi:BPTs}. HII galaxies
(class H) are to the left and below all the curves including the
dashed line in Figure \,\ref{fi:BPTs}a. Composite sources (class C)
lie between the solid and dashed lines in the [NII]/H$\alpha$ diagram
and AGN are above and to the right of the curves in all diagrams. The
AGN population is further divided between Seyfert (class S) or LINER
(class L) using the [SII]/H$\alpha$ and [OI]/H$\alpha$ diagnostic
diagrams and the empirical results of \citet{KEWLEY06} who found two
well defined populations tracing different position angles on these
diagrams. For a significant number of our sources either the [SII] or
[OI] lines were not detected. In order to increase the number of
detections we therefore used the same \emph{2-of-3} criterion for
classification as \citet{YUAN10} whereby if there was no data for a
given line ratio or if one diagram gave a classification inconsistent
with the other two we chose the classification given by the majority
of the diagrams.

Our available optical data allowed for the classification of 66
sources as summarized in column\,10 of Table~\ref{ta:T3stub}. 
The fraction of different classifications agrees well with that found 
amongst sources of similar luminosity by f \citet{YUAN10}.  For
the subset of sources observed by VLBI the compact radio core
detection rates for each optical category are; Seyferts $4/8$ (50\% ),
LINERS 4/6 (66 \%), Composite 3/20 (15\%) and HII 2/15 (14\%). These
VLBI detection fractions are broadly consistent with the idea that
VLBI cores trace AGN activity since the detection fraction is largest
amongst AGN optical types (see \S\ref{optinterp}).


\section{Discussion}\label{se:discussion}
\subsection{Nature of Detected VLBI Sources}\label{se:vlbi_VLA}
The primary objective of our VLBI observations was to detect compact
cores indicating AGN activity. Here we discuss whether our 20 VLBI
detections can be taken as unambiguous signs of AGN activity or
whether they might be confused by the high spatial frequency tail of
arcsecond scale diffuse emission, radio supernovae (SNe), or supernova
remnants (SNRs).

\subsubsection{VLBI Brightness Temperature and Diffuse Starburst
  Emission} \label{compdiffuse}

Could the VLBI detections be due to spatial substructure within the
general ISM starburst-powered radio emission? As noted in
\S\ref{se:jmfit} the flux densities on the Eb--Wb and Eb--Jb1
baselines are usually comparable so  implying a source size less than
about half of the Eb--Jb1 fringe spacing, i.e.\ $\lesssim$10\,mas. A
typical VLBI detection flux density of \about3\,mJy would then imply a
brightness temperature of at least \E{6.2}\,K. Even larger brightness
temperatures are found on the Eb--Ar baseline where sub-milliarcsecond
sizes imply brightness temperatures of $\gtrsim$\E{8}\,K. These high
brightness temperatures are inconsistent with star-formation powered
synchrotron emission which is predicted to have brightness
temperatures of $\lesssim$\E{5}\,K \citep{CONDON91}.

Given that we only have fringe detections on single baselines it is
possible that such detections (especially on the Eb-Jb1 baseline)
could be due to star-formation powered radio emission combined with an
unusual source brightness distribution. One such distribution would be
a thin rectangle with its small dimension $<10$\,mas along the
direction of the fringe spacing; this would however require both
unusual brightness distributions and chance alignments in position
angle in a number of sources to affect our results. A second such
distribution would be one with order 100\% amplitude fluctuations of
\E{5}\,K brightness on scales $<10$\,mas over a region of order
1000\,mas in diameter, giving large r.m.s visibilities of the same
size on Eb-Wb and Eb-Jb1. The ratio of detected visibilities on these
two baselines seems too close to unity to be explained by such a
random model. Additionally such a brightness distribution seems very
unlikely, one would instead expect larger fluctuations as we go to
larger separations within the source giving rise to larger r.m.s
visibilities on the shorter baseline. This is what is in fact observed
in Arp\,220. Taking the data of \citet{PARRA07} and removing the
effect of the compact sources we detect the diffuse emission on Eb-Wb
but not on the Eb-Jb1 baseline. Since Arp\,220 is brighter and more
compact than other COLA sources it seems unlikely that the Eb-Jb1
detections in these can be due to randomly fluctuating diffuse
\E{5}\,K emission.

\subsubsection{Compact Sources in Arp\,220} \label{arp220comp}

For the case of Arp\,220 (15327+2340) most of the flux density
detected in our VLBI snapshot observations clearly comes from non-AGN
activity. This galaxy has been known for many years
\citep{SMITH98,LONSDALE06,PARRA07} to harbor multiple compact radio
sources consistent with a mixed population of luminous radio supernova
remnants (SNRs, interacting with the ISM) and supernovae (SNe,
interacting with progenitor winds).  Arp\,220 is however a very
unusual source within the COLA sample being by far its most IR
luminous object. It follows that a confirmed SNR/SNe origin for the
VLBI snapshot fringes in Arp\,220 does not necessarily imply that the
same explanation is valid for the rest of the VLBI detections. The
diagonal line in Figure~\ref{fi:q_hist} shows the Eb--Wb spectral
luminosity of Arp\,220 scaled linearly with $L_{\rm IR}$ and
demonstrates that if the VLBI contribution from SNR/SNe scales
linearly with star-formation rate then these cannot explain the VLBI
detections. In the next subsections we discuss in more detail whether
SNRs and SNe can in fact explain the VLBI detections.

\subsubsection{Supernova Remnant Origin} \label{SNR} 

\citet{CHOMIUK09}\ find in their study of the radio luminosity
function of SNRs in nearby normal and starburst galaxies a constant
power law exponent and a scaling factor proportional to SFR. They also
find both theoretically and empirically that the luminosity of the
brightest SNR in each galaxy scales linearly with the SFR. Since even
on our shortest VLBI baseline bright SNRs will be separated by several
fringe spacings their expected contribution to the correlated VLBI
flux density will be an incoherent sum of all the SNR flux
densities. This incoherent sum will however, because of the constant
slope SNR luminosity function found by \citet{CHOMIUK09}\, simply be
some multiple of the brightest SNR flux density.  Since this latter
quantity is linearly correlated with SFR we expect the total observed
VLBI signal due to SNRs to also linearly scale with SFR. Since about
half of the flux density seen on the Eb--Wb baseline for Arp\,220
comes from SNR a VLBI contribution which lies a factor of two below
the diagonal line shown in Figure~\ref{fi:q_hist} is therefore
expected; in all cases this lies below the luminosity of the detected
VLBI cores.

In fact estimating the SNR contribution to VLBI flux density in COLA
galaxies by scaling Arp\,220 by $L_{IR}$ as described above probably
results in an overestimate of the SNR contribution for COLA galaxies.
This is because \citet{CHOMIUK09} find that although the SNR
luminosity function in Arp\,220 has the same slope as in other
galaxies the scaling factor between SNR brightness and SFR is larger
(possibly because of the very high gas density in Arp220).  The true
VLBI SNR contribution in COLA galaxies although having the same linear
slope as drawn in Figure\,\ref{fi:q_hist} therefore likely passes
below the Arp\,220 point; because this source is over-luminous in SNR
for its SFR.

\subsubsection{Radio Supernova Origin} \label{SNe}

A difficulty for calculating the possible contribution of radio SNe to
our detected VLBI flux density is that unlike the case for radio SNR
no well defined radio SNe luminosity functions have been published. It
is clear however that only the most luminous Type\,Ic or Type\,IIn
optical classes \citep{CHEVALIER06} would be detectable at the
distance of COLA galaxies. The most powerful Type\,Ic objects are
often associated with $\gamma$-ray bursts and have rapid rise and
decay times. The slightly weaker Type\,IIn SNe are much longer lived
and hence more likely to be detected. Optically about 4\% of
core-collapse SNe are of Type\,IIn \citep{SMARTT09} however even these
show a wide range in radio luminosity. For instance all of the six
IIn's observed by \citet{VANDYK06} were undetected in the radio with
4.8\,GHz spectral luminosity $<$1.4\x\E{19}\,W\,\per{Hz}. The likely
Type\,IIn's in Arp\,220 \citep{PARRA07} have spectral luminosity of
$\about$ 4\x\E{20}\,W\,\per{Hz}. The most radio luminous Type\,IIn so
far detected (2\x\E{21}\,W\,\per{Hz} at 4.8\,GHz) is SN1986J in
NGC\,891 \citep{RUPEN87}. Sources close to this maximum seem however
to be rare. Considering the number of optically identified Type\,IIn's
searched in the radio \citep[see][and references therein]{CHANDRA09}
in normal galactic disks we can estimate that at most one in ten
Type\,IIn's (or 0.4\% of all core collapse SNe) reaches to within a
factor of two of the SN1986J radio luminosity.

Amongst COLA VLBI detections there are six objects other than Arp\,220
with luminosities above that of SN1986J at maximum light (i.e. above
the dashed line in Figure\,\ref{fi:q_hist}). A single RSNe origin
seems to be ruled out for these objects and we are forced to consider
models consisting of two or more extremely luminous objects. Such
SN1986J-like objects stay within a factor of two of their peak
4.8\,GHz luminosity for \about5\,years and host galaxies of these VLBI
detections have total supernova rates of order 0.5\,\per{year}. If
0.4\% of core collapse are SN1986J-like, then we expect 0.01 such
objects on average per COLA galaxy. The probability of getting just
one example of two or more such luminous objects in a galaxy within
our whole sample is thus only 1.5\%.

Figure\,\ref{fi:q_hist} shows 7 VLBI detections at
$\about$\,1\x\E{20}\,W\,\per{Hz} that could potentially be explained
as single SN1986J-like objects observed when above 50\% of maximum
light. Assuming the same radio luminous SNe lifetimes and fractions as
before and typical SFRs of this group of $\about0.25$\,\per{year}
there should only be 0.5 such objects in the whole sample. Finally
there are three VLBI detections at $\about$\,0.6\x\E{20}\,W\,\per{Hz},
it is possible that one of these objects could be a SNe, though
without a detailed radio SN luminosity function it is hard to make any
definite statements.

A general argument that suggests that the vast majority of our VLBI
detections are not explained by SN comes from the fact (see
\S\ref{se:vlbi_det}) that all our VLBI detections are well above our
detection limit. If the origin of the VLBI detections were single or
multiple RSNe such a gap in luminosity between our weakest detection
and the sensitivity limit is hard to explain. A single RSNe spends a
much longer period at low luminosities than at high, and intrinsically
weaker SNe are more common than strong ones, so one would expect many
more detections just above the detection cutoff than are observed.

We conclude that, while one or two of the less luminous detections
could be caused by radio SNe, the vast majority of our VLBI detections
cannot and must instead be AGN powered.

\subsection{Nature of Extended Radio emission}
\label{se:extended}
For most VLBI detected COLA sources (17/20), a large fraction of the
total radio flux density arises from several 100\,pc scale structures
seen in the VLA maps.  As noted in \S\ref{se:sample-q} the total flux
densities of nearly all COLA sources, including both VLBI detections
and non-detections (see \S\ref{se:vlbi-q} and
Figure\,\ref{fi:radio_fir}), closely follows the FIR to radio
correlation at both 1.4 and 4.8\,GHz. This result indicates that most
of the radio emission recovered in the VLA maps is powered by star
formation activity. It is known from observations of the starbursts in
M82 and Arp\,220 that the associated star formation induced compact
emission from SNR and SNe is just a few percent \citep{LONSDALE06} of
the total diffuse component radio emission; hence there is no
contradiction between star-formation being the main power source for
the total radio emission and our conclusion that it does not provide
enough correlated flux density to explain the VLBI detections.

\subsection{Nature of integrated IR emission}
\label{se:natIR}
\begin{figure}[t!]
  \figurenum{7} \centering
  \includegraphics[width=1\hsize]{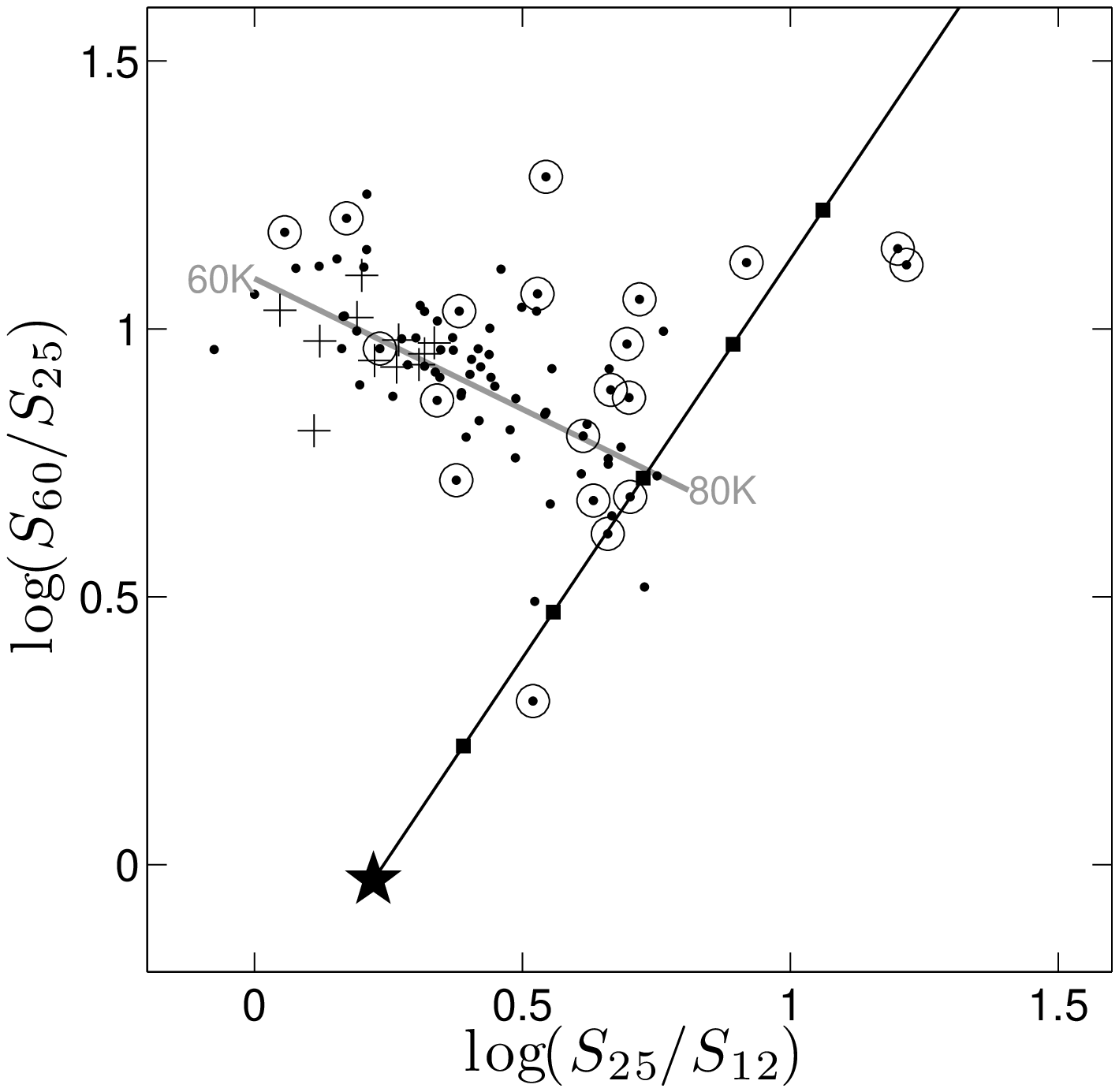}\hfil
  \caption{IR color-color diagram for COLA North sources observed by
    VLBI.  VLBI detections are indicated with open circles and VLA non
    detections as crosses. The gray line is traced by blackbodies of
    temperatures progressing from 60 to 80\,K from the left. The black
    line is a reddening line starting from the average warm Seyfert~1
    nucleus (indicated by a star). Squares over this line are located
    at $\tau_{\rm 25\mu m}$=1, 2, 3, 4 and 5 \citep[see][]{dopita98}.}
  \label{fi:irplots}
\end{figure}

Figure\,\ref{fi:irplots} shows a IR color--color diagram for the 90
sources with VLBI observations in the COLA North sample. Any position
on this diagram can be reached by a linear combination of starburst
temperature, AGN and reddening \citep{KEWLEY00}. It is therefore
impossible to unambiguously estimate the proportions in which these
ingredients are combined. However, most of the sources in the sample
appear to follow the line traced by blackbodies of progressively
increasing temperatures (i.e. the starburst line) with a scatter
apparently increasing with temperature suggesting a larger AGN
contribution to the IR energy budget in warm starbursts. The overall
scatter of the COLA sample is however much smaller than that observed
by \citet{RUSH93} in their sample of optically selected Seyferts. We
therefore argue that for most sources in our sample that their IR
emission is mostly powered by starburst activity with any AGN
contribution being energetically unimportant.

All 10 VLA non-detections (see\,\S\ref{se:VLA_maps}) with VLBI data
are clustered near the cold end of the starburst line (crosses in
Figure\,\ref{fi:irplots}).  As discussed in \S\ref{se:VLA_maps} these
are likely to be large sources whose relatively large FIR luminosity
is not due to concentrated nuclear starburst activity but is instead
caused by the aggregation of normal star forming regions spread over
their large galactic disks.

\subsection{Comparison of optical and radio source
  classifications} \label{optinterp}

As discussed in \S\ref{se:optclass} we find that the fraction of VLBI
detections in AGN-like optical types is substantially higher than in
HII type galaxies, supporting the interpretation that VLBI detections
indicate AGN activity. We find however no one-to-one correspondence
between the optical classification and VLBI detection. The fact that
half of the observed Seyfert sources (4/8) were not detected by VLBI
supports the result of \citet{CORBETT03} who found observationally two
classes of Seyferts with and without detectable VLBI radio cores. This
could simply be an artifact of our limited VLBI sensitivity or it
could mean that there are physically two classes of Seyfert with
different degrees of radio-loudness. The relatively high VLBI
detection rate of LINER sources (4/6) is consistent with models in
which this class is usually AGN powered \citep{KEWLEY06} although it
is clear that in some sources LINER-like spectra can arise in shocks,
i.e. for the case of Arp\,220 \citep{ARRIBAS01}. The relatively low
incidence of VLBI detections in HII galaxies (2/15) suggests that
cases of completely optically obscured AGN are rare.
Combining the VLBI and optical results we find the overall incidence
of AGN activity to be high. If we add the four non-VLBI detected
Seyferts to the 19 VLBI detections (excluding Arp\,220) we get a
minimum fraction of sources containing AGN of 22/90 (25\%).

\begin{figure*}[t!]
  \figurenum{8} \centering
  \includegraphics[width=0.99\hsize]{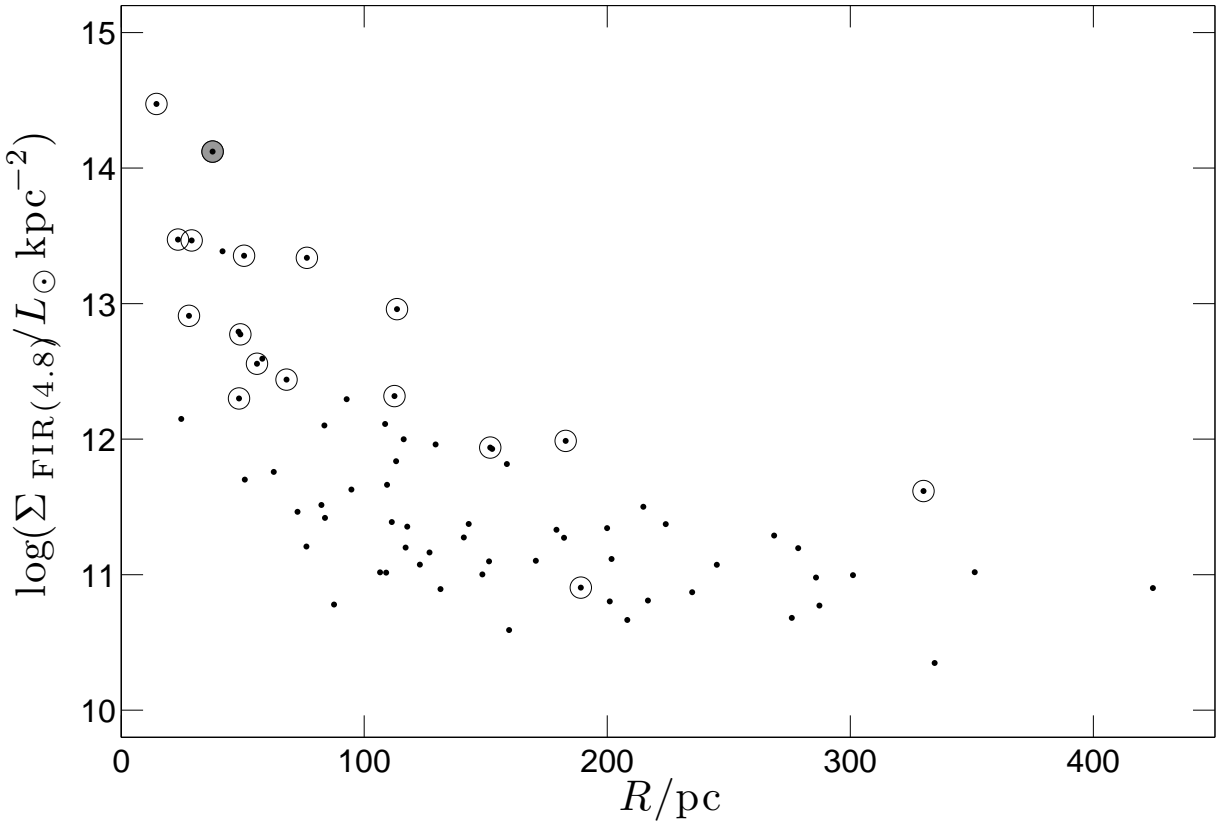}
  \caption{Starburst FIR equivalent brightness $\Sigma_{\rm FIR(4.8)}$
    versus central starburst component radius $R$ for sources both
    imaged by the VLA and observed by VLBI. VLBI detections are
    indicated by a circle with Arp\,220 emphasized as a filled circle.
    This diagram duplicates a similar one in \citet{THOMPSON05} for
    more luminous IR galaxies. In this plot $\Sigma_{\rm
      FIR(4.8)}=L_{\rm FIR(4.8)}/A_{\rm radio}$ where $L_{\rm
      FIR(4.8)}$ is the equivalent starburst FIR spectral luminosity
    estimated from the 4.8\,GHz data and $A_{\rm radio}$ is the area
    of the central starburst source. $L_{\rm FIR(4.8)}$ is calculated
    by transforming the 4.8\,GHz spectral luminosity of the brightest
    VLA component $L_{\rm 4.8 peak}$ to an equivalent FIR spectral
    luminosity using the FIR-radio correlation.
 }.
  \label{fi:todd}
\end{figure*}

\subsection{Implications for the AGN starburst connection}

An interesting correlation found in the present work is that between
the presence of a VLBI detection and of high brightness
arcsecond-scale radio emission (see \S\ref{se:jmfit} and
Figure\,\ref{fi:Tb_hist}). In \S\ref{se:vlbi_VLA} we argued that VLBI
detections are due to AGN activity rather than SNe or SNRs. In
\S\ref{se:extended} we further argued that the extended radio emission
seen with the VLA is due to star formation. Taken together these
results imply a strong connection between AGN activity and the
presence of a compact starburst.


A significant body of evidence in favor of a strong starburst/AGN
connection has been accumulating in recent years.  In nearby AGN
evidence is seen for for nuclear post-starburst stellar populations of
age \about\E{6}--\E{8}\,year \citep{DAVIES07} suggesting a possible
evolutionary connection with nuclear starbursts preceding AGN activity
\citep[see also][]{HECKMAN08}. There are however also many individual
examples of concurrent starburst/AGN activity \citep[e.g.][]{FATHI06}.
The present paper has found evidence linking AGN activity with
concurrent compact starbursts in FIR selected sources and it is
interesting to consider possible physical interpretations of this
result.

The Eddington-limited starbursts discussed by \citet{THOMPSON05},
which can also feed central AGN, are predicted to have central peak
bolometric brightness of \E{13}\,\Lsun\,kpc$^{-2}$ consistent with the
luminous compact starbursts from
\citet{CONDON91}. Figure\,\ref{fi:todd}\ plots the estimated FIR
surface brightness versus fitted size for the arcsecond-scale radio
structures for our COLA sources, matching Figure\,4 of
\citet{THOMPSON05}. We find that our VLBI sources are mostly detected
in sources with bolometric brightness approaching or around
\E{13}\,\Lsun\,kpc$^{-2}$ which coincides with values predicted for
Eddington-limited starbursts.

\citet{THOMPSON05} also argued for two separate modes by which such
compact starbursts can feed a central black hole, supplying either
0.1\Msun\per{year} or 4\Msun\per{year} depending on whether the gas
supply rate at several hundred parsec radius is smaller or larger than
220\Msun\per{year}. The rest of this mass is converted into
stars. Since the star formation rates of COLA sources are $\ll$220
\Msun\per{year} the lower black hole supply rate applies. Assuming an
efficiency of 10\% this supply rate translates into an AGN bolometric
luminosity of 1.5x\E{11}\Lsun\ which is comparable to the IR
luminosities of the COLA galaxies with VLBI detections. Such a high
AGN luminosity, which would mostly emerge in the IR, is inconsistent
with the fact that COLA sources lie on the FIR-radio correlation and
have IR colors typically found in starburst powered sources (see
\ref{se:natIR}).  We can speculate that a constant fraction of gas
supplied to the black hole might be lost to winds or jets formed close
to the black hole event horizon, thus reducing the AGN bolometric
luminosity. The detected core radio emission would be generated at the
base of these jets \citep{FALCKE95}. The starburst regulation of the
gas supply would ensure that these jets had similar mass outflow rates
and mechanical luminosities therefore explaining the relatively narrow
range of radio luminosity of our VLBI detections.
 

The high incidence we find of AGN activity in compact high brightness
starbursts (i.e. VLBI detections in 13/18 sources with specific
star-formation rates $>$\E{12}\,\Lsun\,kpc$^{-2}$, see Figure\,8)
suggests only a short interval between the triggering of such
starbursts and the onset of AGN activity. In contrast \citet{DAVIES07}
argue (based on AGN stellar population studies) that there is a
50--100\,Myr delay between the peak of nuclear starburst activity and
the onset of AGN activity. These studies were however done on AGN
selected sources. AGNs plausibly last much longer after the initial
feeding event than does starburst activity. Possibly this is because
once the central \about1\,pc scale accretion disk is fueled with gas
via a compact starburst long viscous accretion times \citep[up to
\E{9}\,yr.][]{KING07} allow AGN activity to continue long after
starburst activity has faded. Given this timescale difference examples
of currently active nuclear starbursts in AGN selected samples might
well be expected to be rare. Additionally during the initial composite
AGN/starburst phase optical obscuration may be very large making such
objects hard to be optically identified as AGN. Since COLA is FIR
selected and we have used radio observations as the primary means to
detect AGN both of the above biases are avoided allowing the detection
of the initial phases of composite AGN/starburst activity.

Both our own observations and those of \citet{CONDON91} show a strong
correlation between high FIR luminosity and compact high brightness
starbursts. Within the IRAS-BGS sample \citet{CONDON91} found that all
starbursts with $L_{\rm FIR}>\E{11.8}$\,\Lsun\ were in fact of this
high brightness Eddington-limited type. If the correlation between AGN
powered VLBI sources and high brightness shown in
Figure\,\ref{fi:todd} also holds for this more luminous sample then
one would expect all of these to have AGN activity detectable via
VLBI. It can be argued that Arp\,220 is a clear counter-example since
it has high radio brightness and although detected by VLBI this
detection is due to SNe and SNRs rather than an AGN core or
lobes. However as noted by \citet{PARRA07} within the western nucleus
of Arp\,220 there exist several candidate sources for a radio
AGN. Recent millimeter continuum and $^{12}$CO(2--1) spectral line
interferometry results by \citet{DOWNES2007} and \citet{SAKA08}
strongly bolster the case that a powerful AGN does in fact exist in
the western nucleus of Arp\,220.

More generally within snapshot 1.4\,GHz VLBI observations conducted by
\citet{LSL93}, about 50\% of luminous IRAS-BGS sources were detected
on global baselines. These detections had however ambiguous
interpretations being explainable by both AGN and radio SNe/SNR
models. In general, at other wavebands attempts to definitely detect
or reject AGN contributions in these very luminous starbursts have
also often proved inconclusive. If as predicted by the
\citet{THOMPSON05} model for Eddington-limited starbursts the gas
supply to the central black hole is relatively constant for compact
starburst sources we may obtain the same AGN luminosity regardless of
$L_{\rm IR}$. It follows that in more intense higher IR luminosity
starbursts it may be harder to detect the AGN component in the midst
of all the starburst activity. Paradoxically it may be that within the
moderate luminosity COLA sample a strong correlation between intense
starburst activity and radio AGNs is easier to detect than in more
luminous samples. This is firstly because COLA straddles a luminosity
range were there are large numbers of both low brightness and
Eddington-limited starbursts. Secondly, $L_{\rm IR}$ is not high
enough (except for Arp\,220) that there is significant confusion in
the radio between AGN and starburst induced structure.

It is clear that both the VLBI snapshot observations of COLA-N
presented in this paper and the observations of IRAS-BGS sources carried
out by \citet{LSL93} suffer from lack of uv coverage. New long track
VLBI observation  which produce good quality images would help
distinguish star-formation origins of VLBI flux density \citep[by 
for instance seeing multiple compact VLBI components indicating
star-formation, e.g.][]{PARRA07} from AGN origins (core-jet morphology
sources). Observations of half of the VLBI detected COLA-N sources
have already been made using the EVN at 6\,cm wavelength. These data
are presently being reduced.


\section{Conclusions}

The main conclusions of this paper are:
\paragraph{1}
We have made 4.8\,GHz VLA images for 95 of the 110 sources in the COLA
North sample. Of these sources 28, 62 and 5 sources have respectively
Core, Core+Extended and Extended radio morphologies as defined by
\citet{NEFF92}. The 15 non-detected sources are likely extended
(resolved out) galaxies with radio surface brightnesses below our
sensitivity (see~\S\ref{se:VLA_maps}).

\paragraph{2}
We argue that the properties of the 4.8\,GHz radio emission structures
imaged on scales of a few~100\,pc are consistent with being powered by
star formation activity because the recovered flux density in these
structures follows the FIR to radio correlation for star forming
galaxies (see~\S\ref{se:sample-q}).

\paragraph{3}
We detect 22\% (20/90) of the sources observed in the VLBI snapshots
(see~\S\ref{se:vlbi_det}). Assuming that these VLBI detections are
single Gaussian components the simultaneous detection at multiple
baselines in most cases constrains the size of the emitting region to
parsec scales.

\paragraph{4}
The majority of the VLBI detections are well above our detection
threshold (see~Figure~\ref{fi:q_hist}) suggesting a lower limit of
$\about$\E{21}\,W\,\per{Hz} for the luminosity of these compact
sources.

\paragraph{5}
Based on luminosity arguments we can rule out a radio supernova
remnant (SNR) origin for all of the VLBI detections (see \S\ref{SNR}).

\paragraph{6}
We can also rule out a radio supernova (SN) origin for most VLBI
detections unless we assume that the incidence of powerful radio
supernovae is much larger in COLA sources than in most galactic
disks. We estimate that at most one or two of the VLBI detections
could be due to radio SN.

\paragraph{7}
Our VLBI detections are preferentially found in sources with the
highest radio brightness arcsecond-scale radio emission, even after
subtracting off the VLBI cores. Since this arcsec-scale radio emission
is powered by star-formation activity (see point 2 above) we conclude
that there is a relationship between compact starbursts and AGN
activity. This is consistent with the proposal of \citet{THOMPSON05}
that Eddington-limited nuclear starbursts can feed AGN.

\paragraph{8} The high AGN incidence amongst compact starbursts
implies no significant delay between the period of intense nuclear
starburst activity and the onset of AGN activity. If the AGN activity
continues long after the starburst has faded this would explain the
predominance of post starburst stellar populations in AGN sources.

\acknowledgments{\small We would like to thank the anonymous referee
  for his/her very helpful comments on an earlier version of this
  paper. RP acknowledges a Chalmers University PhD student stipend and
  FONDECYT postdoctoral research grant 3085032. JC acknowledges a
  Swedish VR grant. The European VLBI Network is a joint facility of
  European, Chinese, South African and other radio astronomy
  institutes funded by their national research councils. The Arecibo
  Observatory is the principal facility of the National Astronomy and
  Ionosphere Center which is operated by Cornell University under a
  cooperative agreement with the National Science Foundation. The
  National Radio Astronomy Observatory is a facility of the National
  Science Foundation operated under a cooperative agreement by
  Associated Universities, Inc. This research has made use of the
  NASA/IPAC Extragalactic Database (NED) which is operated by the Jet
  Propulsion Laboratory, California Institute of Technology, under
  contract with the National Aeronautics and Space Administration.  }

\bibliography{ms}
\bibliographystyle{apj}

\begin{deluxetable}{ c c c c c c c c c c }
\centering
\tablecolumns{10}
\tablenum{1 (Full Version)}
\tabletypesize{\scriptsize}
\tablewidth{0pt}
\tablecaption{\sc COLA North Sample}
\tablehead{
\colhead{IRAS}&
\colhead{Other}&
\colhead{R.A.}&
\colhead{Decl.}&
\colhead{$V_{\rm Hel}$}&
\colhead{$D_{L}$} & 
\colhead{$\log L_{\rm IR}$}&
\colhead{$\log L_{\rm 1.4}$}&
\colhead{$\log L_{\rm FIR}$}&
\colhead{$q_{1.4}$}
\\
\colhead{Name}&
\colhead{Name}&
\colhead{J2000}&
\colhead{J2000}&
\colhead{[\kms]}&
\colhead{[Mpc]}&
\colhead{[\Lsun]}&
\colhead{[W~\per{Hz}]}&
\colhead{[W~\per{Hz}]}&
\colhead{}
\\
\colhead{(1)}&
\colhead{(2)}&
\colhead{(3)}&
\colhead{(4)}&
\colhead{(5)}&
\colhead{(6)}&
\colhead{(7)}&
\colhead{(8)}&
\colhead{(9)}&
\colhead{(10)}
}
\startdata
00005+2140  	& MRK334  	& 00 03 09.74 	& 21 57 36.46 & 6579  	& 93.4 	& 11.07 	& 22.46 	& 24.74 	& 2.27 \\ 
00073+2538  	& N23  	& 00 09 53.58 	& 25 55 26.40 & 4566  	& 64.7 	& 11.09 	& 22.57 	& 24.82 	& 2.25 \\ 
00506+7248  	&   	& 00 54 04.00 	& 73 05 11.70 & 4706  	& 69.3 	& 11.49 	& 22.82 	& 25.21 	& 2.40 \\ 
00521+2858  	&   	& 00 54 50.27 	& 29 14 47.60 & 4629  	& 65.2 	& 10.88 	& 22.39 	& 24.62 	& 2.23 \\ 
00548+4331  	& N317  	& 00 57 40.58 	& 43 47 32.30 & 5429  	& 77.2 	& 11.16 	& 22.66 	& 24.92 	& 2.26 \\ 
00555+7614  	&   	& \emph{00 59 15} 	& \emph{76 30 52} & 4739  	& 70.1 	& 10.91 	& 22.53 	& 24.66 	& 2.13 \\ 
01503+1227  	&   	& 01 52 59.52 	& 12 42 27.90 & 4558  	& 63.3 	& 10.91 	& 22.36 	& 24.65 	& 2.29 \\ 
01519+3640  	&   	& 01 54 53.93 	& 36 55 04.40 & 5621  	& 79.2 	& 11.02 	& 22.14 	& 24.74 	& 2.60 \\ 
01555+0250  	& ARP126  	& 01 58 05.26 	& 03 05 00.90 & 5431  	& 75.6 	& 10.94 	& 22.47 	& 24.68 	& 2.20 \\ 
01556+2507  	&   	& 01 58 30.63 	& 25 21 36.90 & 4916  	& 68.8 	& 10.99 	& 22.36 	& 24.75 	& 2.39 \\ 
01579+5015  	&   	& 02 01 09.65 	& 50 30 25.50 & 4875  	& 69.5 	& 10.85 	& 22.29 	& 24.55 	& 2.26 \\ 
02071+3857  	& N828  	& 02 10 09.52 	& 39 11 25.30 & 5374  	& 75.8 	& 11.33 	& 22.85 	& 25.07 	& 2.22 \\ 
02080+3725  	& N834  	& \emph{02 11 01} 	& \emph{37 39 58} & 4593  	& 64.7 	& 10.95 	& 22.49 	& 24.69 	& 2.20 \\ 
02152+1418  	& N877  	& 02 17 58.62 	& 14 32 25.90 & 3913  	& 54.2 	& 10.98 	& 22.58 	& 24.74 	& 2.17 \\ 
02208+4744  	&   	& 02 24 08.00 	& 47 58 10.68 & 4679  	& 66.5 	& 11.09 	& 22.51 	& 24.84 	& 2.33 \\ 
02253+1922  	& ARP276  	& \emph{02 28 11} 	& \emph{19 35 45} & 4141  	& 57.5 	& 10.62 	& 22.32 	& 24.37 	& 2.05 \\ 
02345+2053  	& N992  	& 02 37 25.43 	& 21 06 09.41 & 4150  	& 57.6 	& 11.03 	& 22.51 	& 24.76 	& 2.25 \\ 
02346+3412  	&   	& 02 37 40.04 	& 34 25 54.00 & 4915  	& 69.0 	& 10.87 	& 22.04 	& 24.59 	& 2.55 \\ 
02395+3433  	& N1050  	& 02 42 35.62 	& 34 45 48.97 & 3901  	& 54.8 	& 10.76 	& 22.06 	& 24.45 	& 2.39 \\ 
02435+1253  	&   	& 02 46 17.51 	& 13 05 44.56 & 6560  	& 91.8 	& 11.31 	& 22.86 	& 25.08 	& 2.21 \\ 
02438+2122  	&   	& 02 46 39.15 	& 21 35 10.37 & 6987  	& 98.1 	& 11.14 	& 22.31 	& 24.91 	& 2.60 \\ 
02509+1248  	& ARP200  	& 02 53 41.50 	& 13 00 48.67 & 3640  	& 50.4 	& 10.85 	& 22.43 	& 24.59 	& 2.16 \\ 
02511+1238  	&   	& 02 53 50.30 	& 12 50 56.54 & 3568  	& 49.4 	& 10.51 	& 21.86 	& 24.26 	& 2.40 \\ 
02533+0029  	&   	& 02 55 57.25 	& 00 41 33.04 & 4142  	& 57.4 	& 10.85 	& 22.07 	& 24.61 	& 2.55 \\ 
02568+3637  	&   	& 02 59 58.61 	& 36 49 13.79 & 3605  	& 50.8 	& 10.93 	& 22.49 	& 24.62 	& 2.13 \\ 
02572+7002  	&   	& 03 01 59.72 	& 70 14 30.45 & 4890  	& 71.5 	& 10.94 	& 22.42 	& 24.67 	& 2.25 \\ 
03164+4119  	& 3C84  	& 03 19 48.10 	& 41 30 42.00 & 5264  	& 74.5 	& 11.26 	& 24.27 	& 24.75 	& 0.48 \\ 
03251+3958  	&   	& 03 28 27.68 	& 40 09 17.04 & 4262  	& 60.3 	& 10.85 	& 22.27 	& 24.51 	& 2.24 \\ 
03266+4139  	& N1334  	& \emph{03 30 02} 	& \emph{41 49 55} & 4274  	& 60.6 	& 10.78 	& 21.77 	& 24.50 	& 2.73 \\ 
03406+3908  	&   	& 03 43 56.89 	& 39 17 42.50 & 4963  	& 70.2 	& 10.88 	& 22.36 	& 24.61 	& 2.25 \\ 
03449+7252  	&   	& 03 50 36.28 	& 73 01 52.86 & 4312  	& 63.7 	& 10.85 	& 22.13 	& 24.61 	& 2.48 \\ 
03514+1546  	&   	& 03 54 15.96 	& 15 55 43.29 & 6662  	& 93.7 	& 11.18 	& 22.55 	& 24.90 	& 2.35 \\ 
04002+0149  	&   	& 04 02 48.15 	& 01 57 56.65 & 3813  	& 53.1 	& 10.67 	& 22.09 	& 24.41 	& 2.32 \\ 
04007+2201  	&   	& 04 03 44.05 	& 22 09 32.91 & 6261  	& 88.1 	& 11.08 	& 22.51 	& 24.83 	& 2.32 \\ 
04097+0525  	&   	& \emph{04 12 22} 	& \emph{05 32 51} & 5305  	& 74.4 	& 11.19 	& 22.78 	& 24.93 	& 2.15 \\ 
04356+6738  	&   	& 04 40 47.32 	& 67 44 09.32 & 4830  	& 70.8 	& 10.93 	& 22.09 	& 24.56 	& 2.47 \\ 
04435+1822  	&   	& 04 46 29.67 	& 18 27 39.16 & 4615  	& 63.3 	& 10.78 	& 22.21 	& 24.49 	& 2.28 \\ 
04520+0311  	& N1691  	& 04 54 38.38 	& 03 16 04.48 & 4619  	& 65.2 	& 10.93 	& 22.37 	& 24.68 	& 2.31 \\ 
05054+1718  	&   	& 05 08 21.21 	& 17 22 08.29 & 5454  	& 77.4 	& 11.20 	& 22.37 	& 24.97 	& 2.60 \\ 
05091+0508  	& N1819  	& 05 11 46.10 	& 05 12 02.22 & 4470  	& 63.3 	& 10.91 	& 22.31 	& 24.67 	& 2.36 \\ 
05134+5811  	&   	& 05 17 47.58 	& 58 14 19.18 & 5320  	& 77.3 	& 10.96 	& 22.54 	& 24.70 	& 2.16 \\ 
05179+0845  	&   	& 05 20 40.79 	& 08 48 31.20 & 4687  	& 66.6 	& 11.09 	& 22.61 	& 24.86 	& 2.26 \\ 
05336+5407  	&   	& 05 37 46.84 	& 54 09 45.79 & 5392  	& 78.3 	& 10.89 	& 22.19 	& 24.66 	& 2.47 \\ 
05365+6921  	& ARP184  	& 05 42 04.65 	& 69 22 42.35 & 3934  	& 58.6 	& 10.97 	& 22.85 	& 24.72 	& 1.87 \\ 
05405+0035  	&   	& 05 43 05.51 	& 00 37 12.89 & 4318  	& 61.8 	& 10.93 	& 22.01 	& 24.70 	& 2.69 \\ 
05414+5840  	&   	& 05 45 47.97 	& 58 42 04.50 & 4455  	& 65.3 	& 11.27 	& 22.84 	& 25.06 	& 2.21 \\ 
06052+8027  	&   	& \emph{06 14 30} 	& \emph{80 27 00} & 3921  	& 59.3 	& 11.00 	& 22.61 	& 24.76 	& 2.15 \\ 
06140+8220  	&   	& 06 24 57.70 	& 82 19 06.56 & 4317  	& 65.0 	& 10.76 	& 22.57 	& 24.48 	& 1.91 \\ 
06239+7428  	&   	& 06 30 29.73 	& 74 26 31.35 & 5375  	& 79.7 	& 10.98 	& 22.18 	& 24.71 	& 2.53 \\ 
06538+4628  	&   	& 06 57 34.42 	& 46 24 10.51 & 6401  	& 93.7 	& 11.31 	& 22.82 	& 25.05 	& 2.22 \\ 
07062+2041  	&   	& 07 09 12.38 	& 20 36 14.56 & 5227  	& 76.7 	& 11.09 	& 22.66 	& 24.83 	& 2.17 \\ 
07063+2043  	& N2342  	& 07 09 18.08 	& 20 38 09.67 & 5276  	& 77.4 	& 11.30 	& 22.82 	& 25.03 	& 2.20 \\ 
07204+3332  	&   	& 07 23 43.52 	& 33 26 31.38 & 4031  	& 60.1 	& 10.88 	& 22.19 	& 24.59 	& 2.40 \\ 
F07258+3357  	& N2388  	& \emph{07 28 53} 	& \emph{33 49 09} & 4134  	& 61.7 	& 11.24 	& 22.53 	& 25.00 	& 2.47 \\ 
07258+3357  	& N2389  	& 07 29 04.64 	& 33 51 39.10 & 3957  	& 59.1 	& 10.75 	& 22.18 	& 24.52 	& 2.34 \\ 
07329+1149  	&   	& 07 35 43.42 	& 11 42 35.31 & 4873  	& 72.3 	& 11.10 	& 22.46 	& 24.87 	& 2.41 \\ 
07336+3521  	& N2415  	& \emph{07 36 57} 	& \emph{35 14 31} & 3784  	& 56.9 	& 10.92 	& 22.41 	& 24.66 	& 2.25 \\ 
07566+2507  	& N2498  	& 07 59 38.79 	& 24 58 56.73 & 4720  	& 70.7 	& 10.86 	& 22.03 	& 24.60 	& 2.57 \\ 
08287+5246  	&   	& 08 32 28.07 	& 52 36 21.14 & 5094  	& 76.6 	& 10.94 	& 22.06 	& 24.67 	& 2.61 \\ 
08339+6517  	&   	& 08 38 23.42 	& 65 07 14.12 & 5730  	& 85.7 	& 11.11 	& 22.47 	& 24.81 	& 2.34 \\ 
08354+2555  	& ARP243  	& 08 38 24.08 	& 25 45 16.57 & 5549  	& 83.6 	& 11.59 	& 22.90 	& 25.39 	& 2.49 \\ 
08561+0629  	& N2718  	& 08 58 50.47 	& 06 17 34.88 & 3843  	& 59.4 	& 10.66 	& 22.10 	& 24.37 	& 2.27 \\ 
09437+0317  	& I563  	& \emph{09 46 22} 	& \emph{03 03 16} & 6136  	& 94.3 	& 11.26 	& 22.56 	& 24.97 	& 2.41 \\ 
10195+2149  	& N3221  	& \emph{10 22 20} 	& \emph{21 34 10} & 4110  	& 65.2 	& 11.05 	& 22.77 	& 24.80 	& 2.04 \\ 
11413+1103  	& N3839  	& 11 43 54.02 	& 10 47 08.24 & 5916  	& 93.6 	& 11.04 	& 22.70 	& 24.80 	& 2.10 \\ 
11547+2528  	& N3987  	& 11 57 20.92 	& 25 11 42.60 & 4502  	& 72.6 	& 10.98 	& 22.55 	& 24.76 	& 2.21 \\ 
12099+2926  	& N4175  	& 12 12 31.03 	& 29 10 06.28 & 4012  	& 65.5 	& 10.87 	& 22.36 	& 24.63 	& 2.27 \\ 
12159+3005  	& N4253  	& 12 18 26.52 	& 29 48 46.51 & 3876  	& 63.5 	& 10.79 	& 22.26 	& 24.39 	& 2.12 \\ 
12208+0744  	& N4334  	& 12 23 23.87 	& 07 28 23.35 & 4245  	& 69.3 	& 10.88 	& 22.18 	& 24.61 	& 2.43 \\ 
13126+2452  	& IC860  	& 13 15 03.51 	& 24 37 07.78 & 3347  	& 56.4 	& 11.11 	& 22.07 	& 24.92 	& 2.85 \\ 
13183+3423  	& ARP193  	& 13 20 35.33 	& 34 08 22.24 & 6985  	& 109.0 	& 11.70 	& 23.17 	& 25.49 	& 2.32 \\ 
13188+0036  	& N5104  	& 13 21 23.13 	& 00 20 32.90 & 5578  	& 90.2 	& 11.22 	& 22.57 	& 24.99 	& 2.42 \\ 
13238+3611  	& N5149  	& 13 26 09.13 	& 35 56 02.77 & 5652  	& 89.7 	& 11.06 	& 22.41 	& 24.81 	& 2.40 \\ 
13373+0105  	& ARP240  	& \emph{13 39 55} 	& \emph{00 50 02} & 6745  	& 107.0 	& 11.61 	& 22.76 	& 25.33 	& 2.57 \\ 
13470+3530  	&   	& 13 49 15.53 	& 35 15 46.77 & 5032  	& 80.9 	& 11.14 	& 22.87 	& 24.90 	& 2.03 \\ 
14003+3245  	& N5433  	& 14 02 36.04 	& 32 30 36.40 & 4354  	& 71.2 	& 10.99 	& 22.58 	& 24.74 	& 2.16 \\ 
14221+2450  	& N5610  	& 14 24 22.92 	& 24 36 51.15 & 5063  	& 82.0 	& 11.02 	& 22.40 	& 24.75 	& 2.35 \\ 
14280+3126  	& N5653  	& \emph{14 30 10} 	& \emph{31 12 57} & 3562  	& 59.7 	& 11.12 	& 22.49 	& 24.85 	& 2.36 \\ 
15005+8343  	&   	& 14 56 06.77 	& 83 31 22.08 & 3881  	& 60.0 	& 10.78 	& 22.10 	& 24.48 	& 2.38 \\ 
15107+0724  	&   	& 15 13 13.10 	& 07 13 31.85 & 3897  	& 65.0 	& 11.32 	& 22.43 	& 25.15 	& 2.72 \\ 
15276+1309  	& N5936  	& 15 30 00.84 	& 12 59 21.45 & 4004  	& 66.6 	& 11.10 	& 22.88 	& 24.83 	& 1.95 \\ 
15327+2340  	& ARP220  	& 15 34 57.25 	& 23 30 11.41 & 5434  	& 87.3 	& 12.27 	& 23.47 	& 26.07 	& 2.60 \\ 
15437+0234  	& N5990  	& 15 46 16.37 	& 02 24 55.69 & 3839  	& 63.9 	& 11.11 	& 22.51 	& 24.81 	& 2.31 \\ 
16030+2040  	& ARP209  	& 16 05 12.79 	& 20 34 38.38 & 4739  	& 77.1 	& 11.06 	& 22.87 	& 24.79 	& 1.92 \\ 
16478+6303  	& N6247  	& 16 48 19.01 	& 62 58 41.33 & 4480  	& 70.2 	& 10.85 	& 22.37 	& 24.59 	& 2.21 \\ 
16577+5900  	& ARP293  	& 16 58 31.74 	& 58 56 14.65 & 5501  	& 85.1 	& 11.31 	& 23.13 	& 25.10 	& 1.96 \\ 
17180+6039  	& N6361  	& \emph{17 18 41} 	& \emph{60 36 29} & 3881  	& 60.6 	& 10.81 	& 22.30 	& 24.56 	& 2.26 \\ 
17468+1320  	&   	& 17 49 06.20 	& 13 19 55.57 & 4881  	& 77.9 	& 11.01 	& 22.38 	& 24.82 	& 2.44 \\ 
17530+3446  	&   	& \emph{17 54 52} 	& \emph{34 46 34} & 4881  	& 77.0 	& 11.09 	& 22.57 	& 24.82 	& 2.25 \\ 
17548+2401  	&   	& 17 56 56.64 	& 24 01 01.88 & 5944  	& 92.8 	& 11.18 	& 22.74 	& 24.96 	& 2.22 \\ 
18131+6820  	& ARP81  	& 18 12 55.37 	& 68 21 48.24 & 6191  	& 93.6 	& 11.26 	& 22.41 	& 25.00 	& 2.59 \\ 
18145+2205  	&   	& 18 16 40.68 	& 22 06 46.20 & 5599  	& 87.5 	& 11.16 	& 22.46 	& 24.93 	& 2.47 \\ 
18263+2242  	&   	& 18 28 23.92 	& 22 44 11.60 & 4071  	& 65.1 	& 10.77 	& 22.03 	& 24.51 	& 2.48 \\ 
18425+6036  	&   	& 18 43 10.62 	& 60 39 29.55 & 3965  	& 61.9 	& 11.10 	& 22.61 	& 24.86 	& 2.25 \\ 
18495+2334  	&   	& 18 51 37.75 	& 23 38 05.39 & 4559  	& 71.6 	& 10.81 	& 22.12 	& 24.53 	& 2.41 \\ 
19000+4040  	& N6745  	& 19 01 41.24 	& 40 45 04.69 & 4545  	& 70.7 	& 11.01 	& 22.71 	& 24.75 	& 2.05 \\ 
22025+4205  	&   	& 22 04 36.00 	& 42 19 39.46 & 4290  	& 63.5 	& 11.02 	& 22.23 	& 24.80 	& 2.58 \\ 
22171+2908  	& ARP27  	& 22 19 27.73 	& 29 23 44.87 & 4569  	& 66.9 	& 10.94 	& 22.59 	& 24.70 	& 2.11 \\ 
22388+3359  	&   	& 22 41 12.24 	& 34 14 56.51 & 6413  	& 92.8 	& 11.33 	& 22.60 	& 25.09 	& 2.49 \\ 
23007+0836  	& ARP298  	& 23 03 15.62 	& 08 52 26.06 & 4892  	& 70.3 	& 11.63 	& 23.02 	& 25.32 	& 2.30 \\ 
23106+0603  	&   	& 23 13 12.75 	& 06 19 18.11 & 3536  	& 51.0 	& 10.69 	& 21.52 	& 24.29 	& 2.76 \\ 
23157+0618  	&   	& 23 18 16.28 	& 06 35 09.03 & 4966  	& 70.9 	& 11.05 	& 22.50 	& 24.80 	& 2.31 \\ 
23213+0923  	&   	& 23 23 54.07 	& 09 40 02.10 & 3559  	& 51.1 	& 10.58 	& 21.72 	& 24.33 	& 2.60 \\ 
23262+0314  	& ARP216  	& 23 28 46.67 	& 03 30 41.00 & 5138  	& 73.3 	& 11.13 	& 22.55 	& 24.83 	& 2.28 \\ 
23414+0014  	& N7738  	& 23 44 02.04 	& 00 30 59.84 & 6762  	& 96.3 	& 11.11 	& 22.57 	& 24.86 	& 2.29 \\ 
23445+2911  	& ARP86  	& \emph{23 47 02} 	& \emph{29 28 23} & 4845  	& 69.1 	& 10.95 	& 22.32 	& 24.71 	& 2.39 \\ 
23485+1952  	& N7769  	& 23 51 03.98 	& 20 09 01.52 & 4211  	& 59.9 	& 10.99 	& 22.41 	& 24.77 	& 2.36 \\ 
23488+1949  	& N7771  	& 23 51 24.85 	& 20 06 42.30 & 4256  	& 58.7 	& 11.31 	& 22.77 	& 25.08 	& 2.32 \\ 
23488+2018  	& MRK331  	& 23 51 26.74 	& 20 35 10.33 & 5541  	& 78.7 	& 11.49 	& 22.72 	& 25.24 	& 2.52 \\ 
23591+2312  	& Taffy  	& 00 01 41.91 	& 23 29 45.00 & 4353  	& 61.8 	& 10.91 	& 22.79 	& 24.66 	& 1.88 \\ 
\enddata
\label{ta:T1full}

\begin{minipage}[t]{0.90\hsize}
{\sc Columns---}
(1): IRAS name (from Point Source Calalog, excepting  F07258+3357 which is 
taken from the Faint Source Catlog).
(2):  Other name.
(3)-(4): Right Ascencion and Declination (from VLA observations, unless undetected in which case
IRAS positions are given in italics).  VLA positions are conservatively estimated to be accurate 
to 0.3$''$ and IRAS positions to 20$''$. 
(5): Heliocentric velocity (cz) from NED.
(6): Luminosity distance (see Section 2).
(7): FIR luminosity (see Section 3.3).
(8): 1.4~GHz luminosity calculated from the NVSS catalog \citep{NVSS}.
(9): FIR spectral luminosity (see Section 3,3).
(10)): $q_{\nu}=\log(L_{\rm FIR}/L_{1.4})$.
\end{minipage}

\end{deluxetable}

\begin{deluxetable}{ c c c c c c c c c c }
\centering
\tablecolumns{10}
\tablenum{3 (Full Version)}
\tabletypesize{\scriptsize}
\tablewidth{0pt}
\tablecaption{\sc Radio and optical observations and results}
\tablehead{
\colhead{IRAS}&
\colhead{VLA }&
\colhead{VLA }&
\colhead{$S_{\rm 4.8~total}$}&
\colhead{$q_{4.8}$ } &
\colhead{$S_{\rm 4.8~comp}$}&
\colhead{$\theta_{M}\x\theta_{m}$}&
\colhead{$\log T_b$}&
\colhead{VLBI} &
\colhead{Optical}
\\
\colhead{Name}&
\colhead{Array}&
\colhead{Morph}&
\colhead{(mJy)}&
\colhead{}&
\colhead{(mJy)}&
\colhead{[\arcsec\x\arcsec]}&
\colhead{[K]}&
\colhead{Epoch} &
\colhead{Class}
\\
\colhead{(1)}&
\colhead{(2)}&
\colhead{(3)}&
\colhead{(4)}&
\colhead{(5)}&
\colhead{(6)}&
\colhead{(7)}&
\colhead{(8)}&
\colhead{(9)}&
\colhead{(10)}
\\
}
\startdata
00005+2140 	& BnA--A 	& CE 	& 5.66 	& 2.97 	& 0.00 	& -- 	& -- 	& 1*  	& S  \\ 
00073+2538 	& BnA--A 	& C 	& 18.40 	& 2.85 	& 1.23 	& 0.45\x0.22 	& 2.63 	& 1   	& C  \\ 
00506+7248 	& A 	& CE 	& 31.60 	& 2.95 	& 13.60 	& 1.27\x0.66 	& 2.75 	& 2   	& --  \\ 
00521+2858 	& BnA 	& E 	& 6.07 	& 3.13 	& 3.40 	& 3.15\x0.91 	& 1.61 	& 1   	& --  \\ 
00548+4331 	& A 	& CE 	& 13.21 	& 2.95 	& 13.11 	& 1.51\x0.41 	& 2.87 	& 2   	& --  \\ 
00555+7614 	& A 	& U 	& -- 	& -- 	& -- 	& -- 	& -- 	& 2   	& --  \\ 
01503+1227 	& BnA 	& CE 	& 6.23 	& 3.18 	& 5.25 	& 1.58\x1.03 	& 2.05 	& 1   	& C  \\ 
01519+3640 	& A 	& CE 	& 4.38 	& 3.23 	& 3.92 	& 1.13\x0.74 	& 2.21 	& 1-2 	& C  \\ 
01555+0250 	& BnA 	& CE 	& 8.23 	& 2.92 	& 1.85 	& 2.04\x1.06 	& 1.47 	& --  	& L  \\ 
01556+2507 	& BnA 	& CE 	& 11.16 	& 2.95 	& 7.37 	& 1.32\x1.02 	& 2.28 	& 1   	& H  \\ 
01579+5015 	& A 	& CE 	& 2.49 	& 3.39 	& 1.82 	& 0.59\x0.38 	& 2.45 	& 2   	& L  \\ 
02071+3857 	& A 	& E 	& 12.20 	& 3.15 	& 0.93 	& 0.85\x0.56 	& 1.83 	& 2   	& --  \\ 
02080+3725 	& A 	& U 	& -- 	& -- 	& -- 	& -- 	& -- 	& 1-2 	& H  \\ 
02152+1418 	& BnA 	& CE 	& 2.39 	& 3.82 	& 1.34 	& 1.49\x0.94 	& 1.52 	& 1-2 	& --  \\ 
02208+4744 	& A 	& CE 	& 10.92 	& 3.08 	& 10.61 	& 1.78\x1.02 	& 2.31 	& 2   	& --  \\ 
02253+1922 	& BnA 	& U 	& -- 	& -- 	& -- 	& -- 	& -- 	& 1-2 	& H  \\ 
02345+2053 	& BnA 	& CE 	& 14.69 	& 2.99 	& 3.61 	& 1.90\x1.20 	& 1.74 	& 1-2 	& C  \\ 
02346+3412 	& A 	& CE 	& 4.88 	& 3.15 	& 4.06 	& 2.16\x1.28 	& 1.71 	& 1-2 	& H  \\ 
02395+3433 	& A 	& C 	& 3.62 	& 3.34 	& 2.98 	& 0.53\x0.40 	& 2.69 	& 2   	& C  \\ 
02435+1253 	& BnA 	& CE 	& 20.60 	& 2.76 	& 14.98 	& 4.21\x1.44 	& 1.93 	& 1-2 	& S  \\ 
02438+2122 	& BnA 	& C 	& 17.44 	& 2.61 	& 17.40 	& 0.28\x0.10 	& 4.33 	& 2   	& L  \\ 
02509+1248 	& BnA 	& CE 	& 11.38 	& 3.05 	& 3.16 	& 2.11\x1.32 	& 1.59 	& --  	& C  \\ 
02511+1238 	& BnA 	& CE 	& 5.34 	& 3.07 	& 3.26 	& 1.91\x1.51 	& 1.59 	& 1-2 	& H  \\ 
02533+0029 	& BnA-A 	& CE 	& 4.65 	& 3.35 	& 7.06 	& 0.60\x0.54 	& 2.88 	& --  	& H  \\ 
02568+3637 	& A 	& CE 	& 28.16 	& 2.68 	& 30.65 	& 2.10\x0.69 	& 2.86 	& 2*  	& S  \\ 
02572+7002 	& A 	& CE 	& 3.78 	& 3.30 	& 1.82 	& 0.81\x0.62 	& 2.10 	& 2   	& --  \\ 
03164+4119 	& A 	& C 	& 49400.00 	& -0.76 	& 0.00 	& -- 	& -- 	& 1-2* 	&  L \\ 
03251+3958 	& A 	& CE 	& 4.79 	& 3.19 	& 1.02 	& 1.07\x0.24 	& 2.14 	& 2   	& H  \\ 
03266+4139 	& A 	& U 	& -- 	& -- 	& -- 	& -- 	& -- 	& --  	& --  \\ 
03406+3908 	& A 	& C 	& 0.40 	& 4.24 	& 0.41 	& 0.26\x0.00 	& 3.04 	& 2   	&  S  \\ 
03449+7252 	& A 	& CE 	& 4.98 	& 3.22 	& 2.12 	& 0.81\x0.67 	& 2.13 	& 2   	& --  \\ 
03514+1546 	& BnA 	& CE 	& 4.07 	& 3.27 	& 3.51 	& 1.56\x0.93 	& 1.92 	& 1   	& --  \\ 
04002+0149 	& BnA 	& CE 	& 5.99 	& 3.10 	& 7.80 	& 4.68\x1.78 	& 1.51 	& --  	& L  \\ 
04007+2201 	& BnA 	& CE 	& 2.98 	& 3.39 	& 1.31 	& 1.02\x0.80 	& 1.74 	& 1   	& C  \\ 
04097+0525 	& A 	& U 	& -- 	& -- 	& -- 	& -- 	& -- 	& --  	& H  \\ 
04356+6738 	& A 	& CE 	& 6.07 	& 3.00 	& 5.61 	& 1.53\x1.25 	& 2.01 	& 2   	& --  \\ 
04435+1822 	& BnA--A 	& C 	& 7.90 	& 2.91 	& 4.64 	& 1.88\x0.05 	& 3.23 	& 1-2* 	& H  \\ 
04520+0311 	& BnA 	& CE 	& 11.04 	& 2.93 	& 4.42 	& 3.42\x1.25 	& 1.55 	& --  	& C  \\ 
05054+1718 	& BnA-A 	& C 	& 9.45 	& 3.14 	& 7.43 	& 0.33\x0.25 	& 3.49 	& 1*  	&  S   \\ 
05091+0508 	& BnA 	& CER 	& 13.06 	& 2.87 	& 0.43 	& 1.15\x0.67 	& 1.29 	& --  	& C  \\ 
05134+5811 	& A 	& CE 	& 1.62 	& 3.64 	& 0.57 	& 0.27\x0.06 	& 3.09 	& 2   	& --  \\ 
05179+0845 	& BnA 	& CE 	& 3.68 	& 3.57 	& 1.30 	& 1.68\x0.97 	& 1.44 	& --  	& H  \\ 
05336+5407 	& A 	& CE 	& 5.48 	& 3.06 	& 3.56 	& 0.91\x0.34 	& 2.60 	& 2   	& --  \\ 
05365+6921 	& A 	& C 	& 1.36 	& 3.97 	& 1.45 	& 0.06\x0.00 	& 4.22 	& 2   	& --  \\ 
05405+0035 	& BnA 	& CE 	& 4.99 	& 3.34 	& 5.53 	& 1.42\x0.87 	& 2.19 	& --  	& H  \\ 
05414+5840 	& A 	& CE 	& 21.80 	& 3.01 	& 15.75 	& 0.83\x0.39 	& 3.23 	& 2   	& --  \\ 
06052+8027 	& A 	& U 	& -- 	& -- 	& -- 	& -- 	& -- 	& 2   	& --  \\ 
06140+8220 	& A 	& C 	& 57.69 	& 2.02 	& 58.72 	& 0.20\x0.04 	& 5.41 	& 2*  	& --  \\ 
06239+7428 	& A 	& C 	& 1.69 	& 3.60 	& 0.00 	& -- 	& -- 	& 2*  	& --  \\ 
06538+4628 	& A 	& CE 	& 6.11 	& 3.24 	& 3.97 	& 0.57\x0.40 	& 2.78 	& 2   	& --  \\ 
07062+2041 	& BnA 	& CE 	& 6.77 	& 3.15 	& 2.48 	& 0.97\x0.81 	& 2.04 	& 1   	& H  \\ 
07063+2043 	& BnA-A 	& CE 	& 7.07 	& 3.32 	& 4.55 	& 1.18\x0.72 	& 2.27 	& 1   	& --  \\ 
07204+3332 	& A 	& CE 	& 7.77 	& 3.06 	& 1.68 	& 0.56\x0.42 	& 2.39 	& 1-2 	& --  \\ 
F07258+3357 	& A 	& CE 	& 12.95 	& 3.23 	& 13.26 	& 0.56\x0.35 	& 3.37 	& 1-2* 	& H  \\ 
07258+3357 	& A 	& U 	& -- 	& -- 	& -- 	& -- 	& -- 	& 1   	& --  \\ 
07329+1149 	& BnA 	& C 	& 10.62 	& 3.05 	& 10.23 	& 0.75\x0.55 	& 2.93 	& 1   	& H  \\ 
07336+3521 	& A 	& U 	& -- 	& -- 	& -- 	& -- 	& -- 	& 1   	& --  \\ 
07566+2507 	& BnA 	& CE 	& 3.57 	& 3.27 	& 3.25 	& 1.65\x1.07 	& 1.80 	& 1   	& C  \\ 
08287+5246 	& A 	& CE 	& 4.58 	& 3.17 	& 1.21 	& 0.91\x0.45 	& 2.01 	& 2   	& --  \\ 
08339+6517 	& A 	& CE 	& 6.04 	& 3.08 	& 1.69 	& 0.62\x0.48 	& 2.29 	& 2   	& H  \\ 
08354+2555 	& BnA--A 	& C 	& 71.49 	& 2.61 	& 72.08 	& 0.44\x0.30 	& 4.28 	& 1*  	& -  \\ 
08561+0629 	& BnA 	& CE 	& 2.26 	& 3.39 	& 1.93 	& 2.01\x1.32 	& 1.40 	& --  	& H  \\ 
09437+0317 	& BnA 	& U 	& -- 	& -- 	& -- 	& -- 	& -- 	& --  	& --  \\ 
10195+2149 	& BnA 	& U 	& -- 	& -- 	& -- 	& -- 	& -- 	& 1   	&  --   \\ 
11413+1103 	& BnA 	& CE 	& 8.02 	& 2.88 	& 1.13 	& 2.58\x0.78 	& 1.29 	& 1   	& H  \\ 
11547+2528 	& BnA 	& E 	& 8.68 	& 3.03 	& 9.68 	& 4.81\x0.78 	& 1.95 	& 1   	& --  \\ 
12099+2926 	& BnA 	& E 	& 11.89 	& 2.84 	& 13.26 	& 4.43\x1.53 	& 1.83 	& 1   	& --  \\ 
12159+3005 	& BnA 	& C 	& 14.16 	& 2.55 	& 14.24 	& 0.36\x0.26 	& 3.72 	& 1   	& S  \\ 
12208+0744 	& BnA 	& CE 	& 10.08 	& 2.85 	& 2.37 	& 1.37\x0.56 	& 2.03 	& 2   	& H  \\ 
13126+2452 	& A 	& C 	& 19.96 	& 3.03 	& 20.23 	& 0.20\x0.14 	& 4.40 	& 1*  	& --   \\ 
13183+3423 	& A 	& CE 	& 49.12 	& 2.65 	& 39.13 	& 0.58\x0.29 	& 3.91 	& 1-2* 	& C  \\ 
13188+0036 	& BnA 	& C 	& 7.64 	& 3.11 	& 7.13 	& 1.15\x0.78 	& 2.44 	& 2   	& C  \\ 
13238+3611 	& A 	& CE 	& 6.36 	& 3.02 	& 1.84 	& 0.75\x0.52 	& 2.21 	& 1-2 	& --  \\ 
13373+0105 	& BnA 	& U 	& -- 	& -- 	& -- 	& -- 	& -- 	& --  	& --  \\ 
13470+3530 	& A 	& C 	& 3.87 	& 3.42 	& 3.82 	& 0.27\x0.07 	& 3.84 	& 1-2* 	& --  \\ 
14003+3245 	& A 	& CE 	& 9.48 	& 2.98 	& 2.37 	& 0.77\x0.51 	& 2.32 	& 1-2 	& C  \\ 
14221+2450 	& A 	& C 	& 8.40 	& 2.92 	& 5.37 	& 0.35\x0.29 	& 3.26 	& --  	& C  \\ 
14280+3126 	& A 	& U 	& -- 	& -- 	& -- 	& -- 	& -- 	& 1-2 	& H  \\ 
15005+8343 	& A 	& CE 	& 5.28 	& 3.13 	& 1.30 	& 1.04\x0.49 	& 1.95 	& 2   	& --  \\ 
15107+0724 	& BnA 	& C 	& 31.69 	& 2.94 	& 32.34 	& 0.30\x0.26 	& 4.16 	& --  	&   C  \\ 
15276+1309 	& BnA--A 	& CE 	& 11.23 	& 3.06 	& 3.40 	& 0.64\x0.51 	& 2.56 	& 1   	& C  \\ 
15327+2340 	& BnA 	& CE 	& 183.13 	& 2.85 	& 97.51 	& 0.21\x0.14 	& 5.06 	& 1*  	& L  \\ 
15437+0234 	& BnA 	& CE 	& 6.69 	& 3.30 	& 6.51 	& 1.99\x1.67 	& 1.83 	& --  	& S  \\ 
16030+2040 	& BnA 	& CE 	& 14.59 	& 2.78 	& 1.23 	& 0.61\x0.31 	& 2.35 	& 1   	& H  \\ 
16478+6303 	& A 	& E 	& 41.48 	& 2.20 	& 36.32 	& 2.48\x1.43 	& 2.55 	& 2*  	& --  \\ 
16577+5900 	& A 	& C 	& 15.24 	& 2.98 	& 7.80 	& 0.25\x0.21 	& 3.71 	& 2*  	&  C  \\ 
17180+6039 	& A 	& U 	& -- 	& -- 	& -- 	& -- 	& -- 	& 2   	& --  \\ 
17468+1320 	& BnA--A 	& C 	& 24.43 	& 2.57 	& 16.07 	& 0.17\x0.13 	& 4.40 	& 1-2* 	& --  \\ 
17530+3446 	& A 	& U 	& -- 	& -- 	& -- 	& -- 	& -- 	& 1-2 	& --  \\ 
17548+2401 	& BnA--A 	& C 	& 16.19 	& 2.73 	& 12.08 	& 0.70\x0.33 	& 3.26 	& 1*  	& L  \\ 
18131+6820 	& A 	& CE 	& 6.37 	& 3.18 	& 6.34 	& 2.19\x0.59 	& 2.23 	& 2   	& C  \\ 
18145+2205 	& BnA 	& CE 	& 6.80 	& 3.13 	& 6.30 	& 2.67\x0.60 	& 2.13 	& 1   	& --  \\ 
18263+2242 	& BnA--A 	& CE 	& 3.73 	& 3.23 	& 1.15 	& 0.87\x0.52 	& 1.94 	& 1   	& C  \\ 
18425+6036 	& A 	& CE 	& 15.37 	& 3.01 	& 2.29 	& 0.13\x0.00 	& 4.09 	& 2   	& C  \\ 
18495+2334 	& BnA 	& CE 	& 6.43 	& 2.93 	& 6.95 	& 1.76\x1.61 	& 1.93 	& 1   	& C  \\ 
19000+4040 	& A 	& CE 	& 2.77 	& 3.53 	& 1.38 	& 1.25\x0.44 	& 1.94 	& --  	& H  \\ 
22025+4205 	& A 	& C 	& 15.72 	& 2.92 	& 13.01 	& 0.48\x0.28 	& 3.53 	& 2   	&  --  \\ 
22171+2908 	& BnA 	& C 	& 6.40 	& 3.17 	& 4.28 	& 1.33\x0.55 	& 2.31 	& 1   	& --  \\ 
22388+3359 	& A 	& CE 	& 10.86 	& 3.04 	& 7.03 	& 0.51\x0.42 	& 3.06 	& 1-2 	& C  \\ 
23007+0836 	& BnA--A 	& C 	& 63.66 	& 2.75 	& 46.20 	& 0.33\x0.25 	& 4.29 	& 2*  	& S  \\ 
23106+0603 	& BnA--A 	& CE 	& 6.15 	& 3.00 	& 3.79 	& 0.45\x0.22 	& 3.12 	& --  	& H-  \\ 
23157+0618 	& BnA--A 	& C 	& 12.04 	& 2.94 	& 12.05 	& 1.02\x0.52 	& 2.90 	& 2   	& S  \\ 
23213+0923 	& BnA 	& CE 	& 5.34 	& 3.10 	& 2.27 	& 1.73\x1.36 	& 1.52 	& --  	& H  \\ 
23262+0314 	& BnA 	& CE 	& 7.99 	& 3.12 	& 1.52 	& 1.38\x0.58 	& 1.82 	& --  	& S  \\ 
23414+0014 	& BnA--A 	& CE 	& 10.86 	& 2.78 	& 3.77 	& 0.51\x0.23 	& 3.05 	& 2   	& --  \\ 
23445+2911 	& BnA 	& U 	& -- 	& -- 	& -- 	& -- 	& -- 	& 1   	& --  \\ 
23485+1952 	& BnA 	& C 	& 0.54 	& 4.40 	& 0.51 	& 0.86\x0.40 	& 1.71 	& 1   	& C  \\ 
23488+1949 	& BnA 	& CER 	& 23.88 	& 3.09 	& 3.30 	& 1.67\x1.00 	& 1.84 	& 1*  	& L  \\ 
23488+2018 	& BnA-A 	& C 	& 31.23 	& 2.87 	& 20.77 	& 1.17\x0.73 	& 2.93 	& 1-2* 	& C  \\ 
23591+2312 	& BnA 	& C 	& 14.09 	& 2.85 	& 2.30 	& 1.63\x0.57 	& 1.93 	& 1   	& C  \\ 
\enddata
\label{ta:T3full}

\begin{minipage}[t]{0.90\hsize}
{\sc Columns---}
(1): IRAS Name
(2): VLA array(s) used
(3): Morphology class of VLA 4.8GHz image. C-Compact,  E-Extended,  CE-Compact plus Extended, CER-Compact plus extended with ring, U-Undetected.
(4): Recovered 4.8GHz flux density in VLA image.
(5). Log of ratio of 4.8GHz recovered VLA flux density  to FIR flux density.
(6): Total flux density of gaussian fitted to brightest feature in VLA image (after removing VLBI component, hence 
 zero indicates that brightest feature was dominated by the VLBI component).  
  (7): Fitted gaussian major and minor axes to peak VLA feature (FWHM arcsec), Note 
  for  IRAS15327+2340 (Arp~220)  these are for the western nucleus as taken from the literature (see Section  \ref{se:jmfit}).
(7): Fitted gaussian major and minor axes to peak VLA feature (FWHM arcsec).
(8): Peak  4.8GHz brightness temperature of gaussian component  (see Section  \ref{se:jmfit}). Note if formally unresolved along one axis 
a 50mas FWHM was assumed along that axis).
(9): VLBI Epoch(s) observed (asterisk indicates VLBI detection, see Table~\ref{ta:VLBI_summary} for details).
(10): Optical Class  (see Section \ref{se:optclass})  with {\tt H}=HII, {\tt S}=Seyfert, {\tt
    L}=LINER and {\tt C}=Composite.
\end{minipage}

\end{deluxetable}

\begin{figure*}[ht!]
\figurenum{3.2}
\centering
\includegraphics[width=0.9\hsize]{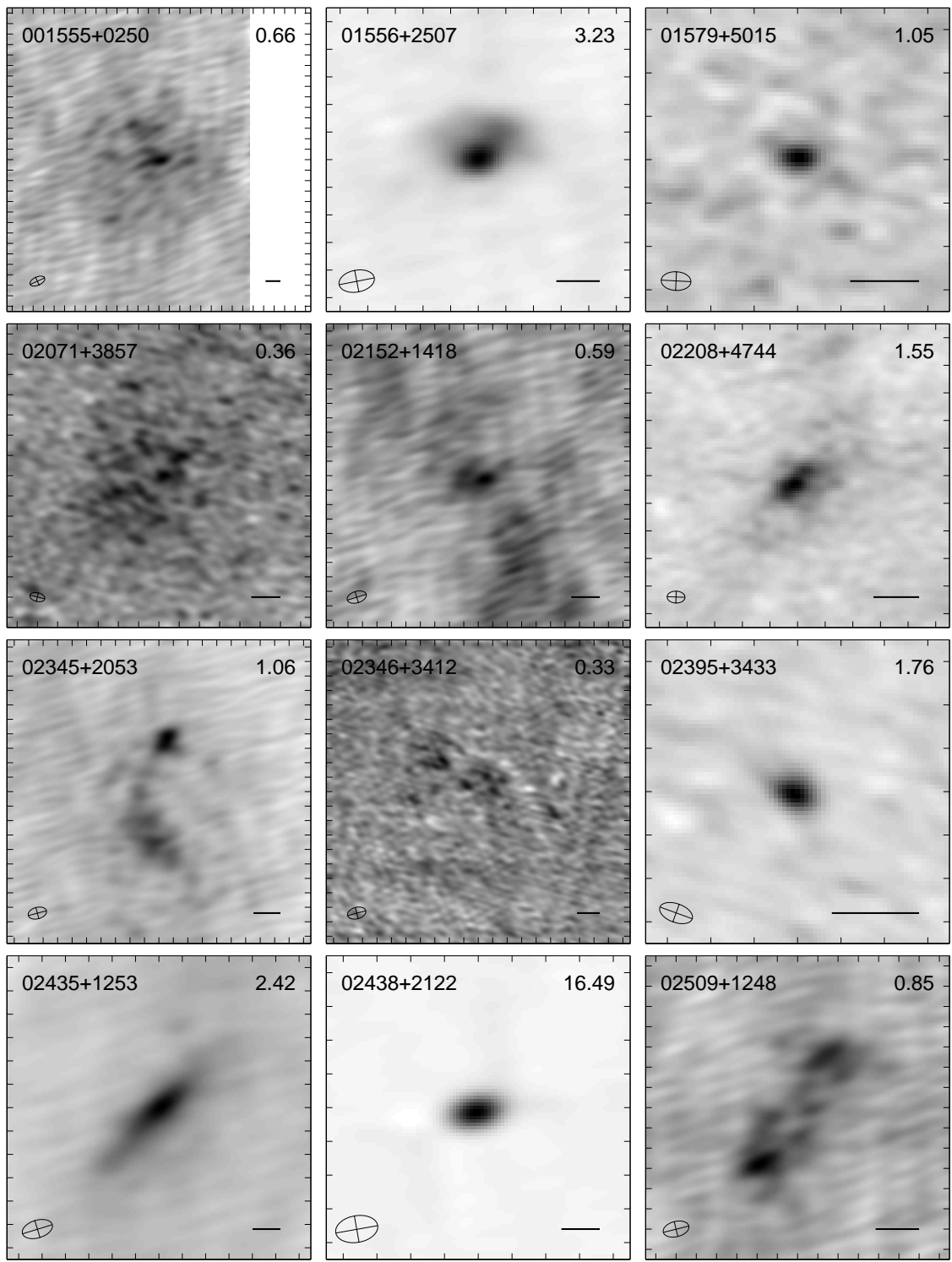}
\caption{See caption of Figure~\ref{fi:nondet1}}
\label{fi:nondet2}
\end{figure*}

\begin{figure*}[ht!]
\figurenum{3.3}
\centering
\includegraphics[width=0.9\hsize]{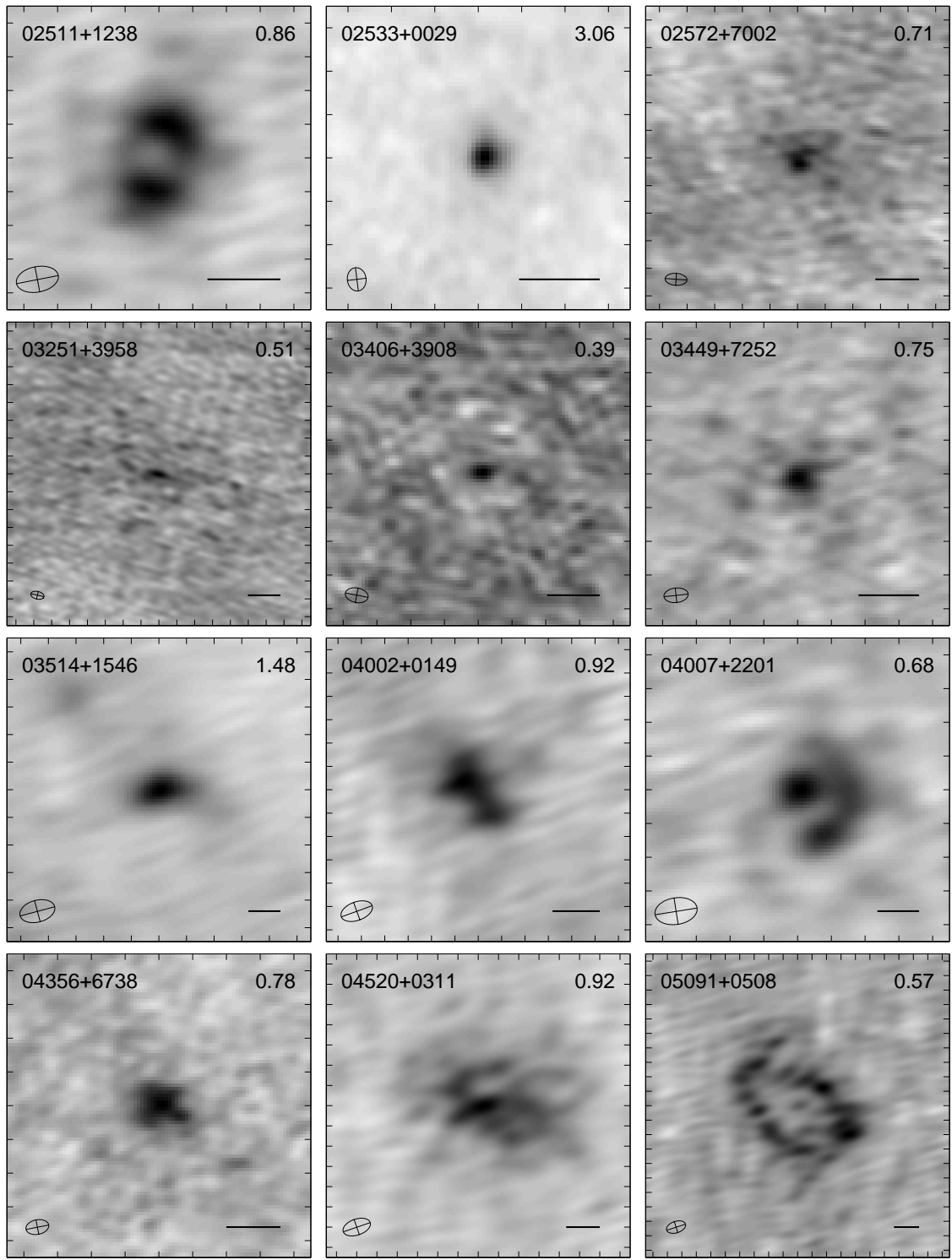}
\caption{See caption of Figure~\ref{fi:nondet1}}
\label{fi:nondet3}
\end{figure*}

\begin{figure*}[ht!]
\figurenum{3.4}
\centering
\includegraphics[width=0.9\hsize]{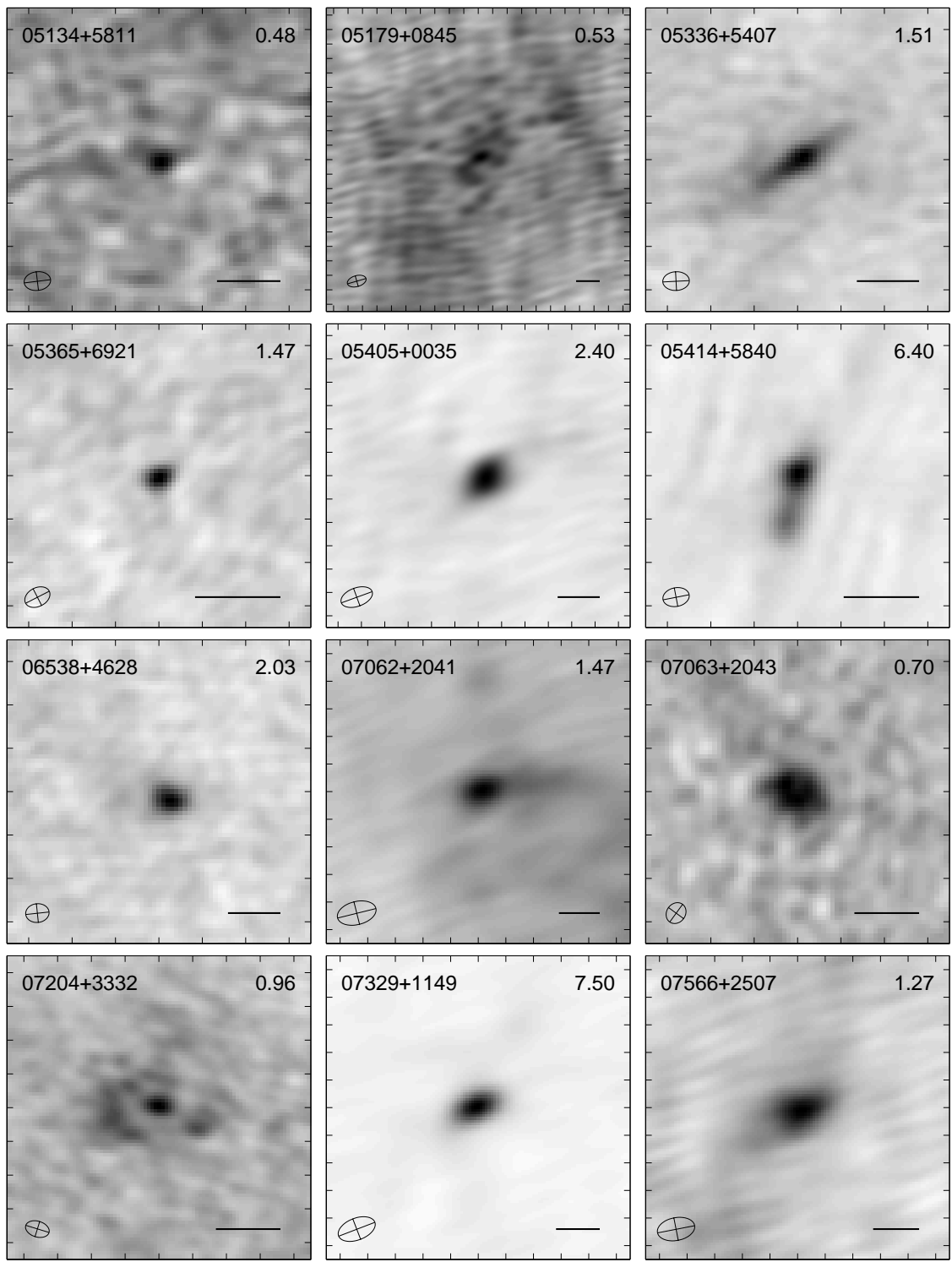}
\caption{See caption of Figure~\ref{fi:nondet1}}
\label{fi:nondet4}
\end{figure*}

\begin{figure*}[ht!]
\figurenum{3.5}
\centering
\includegraphics[width=0.9\hsize]{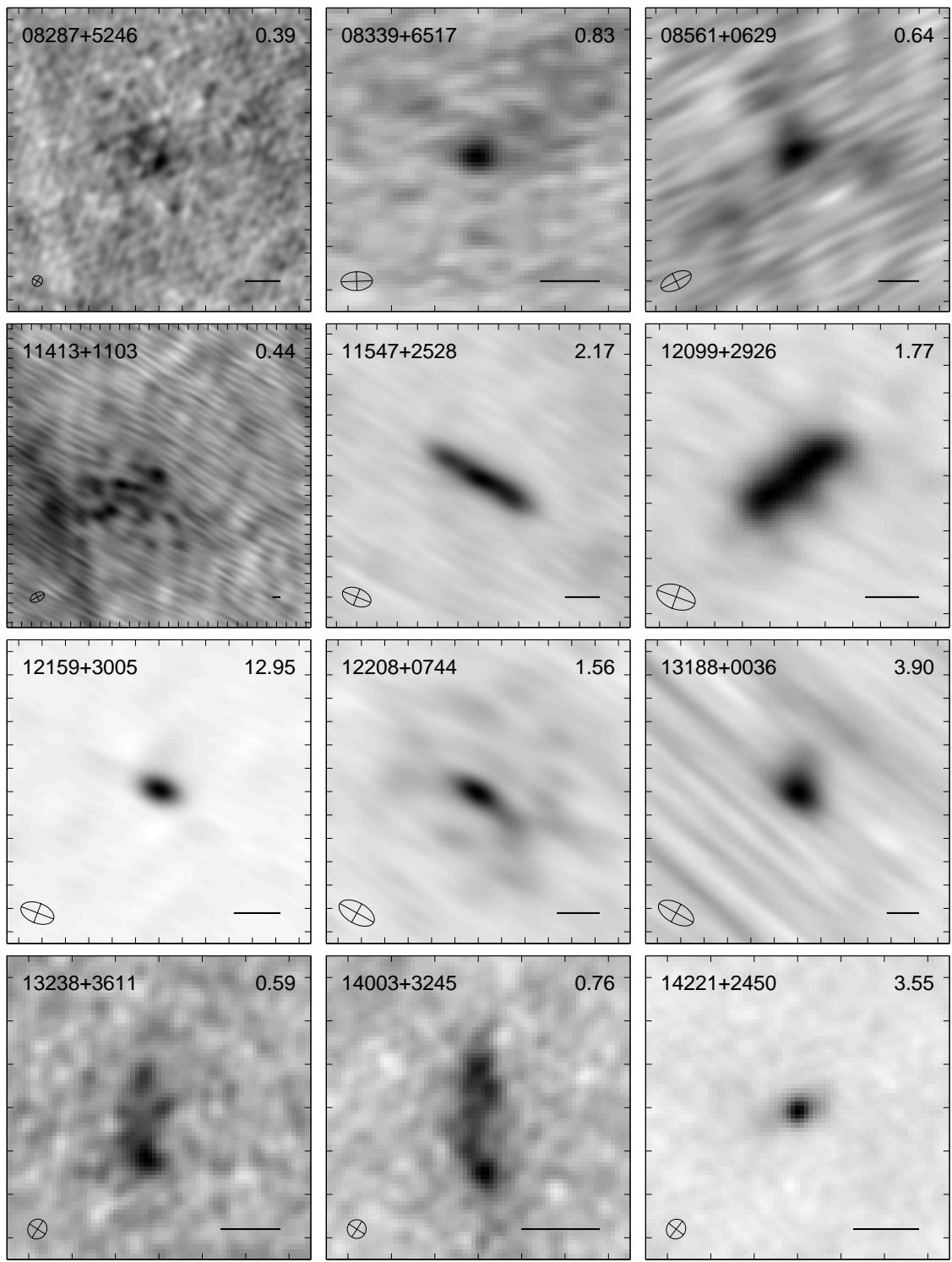}
\caption{See caption of Figure~\ref{fi:nondet1}}
\label{fi:nondet5}
\end{figure*}

\begin{figure*}[ht!]
\figurenum{3.6}
\centering
\includegraphics[width=0.9\hsize]{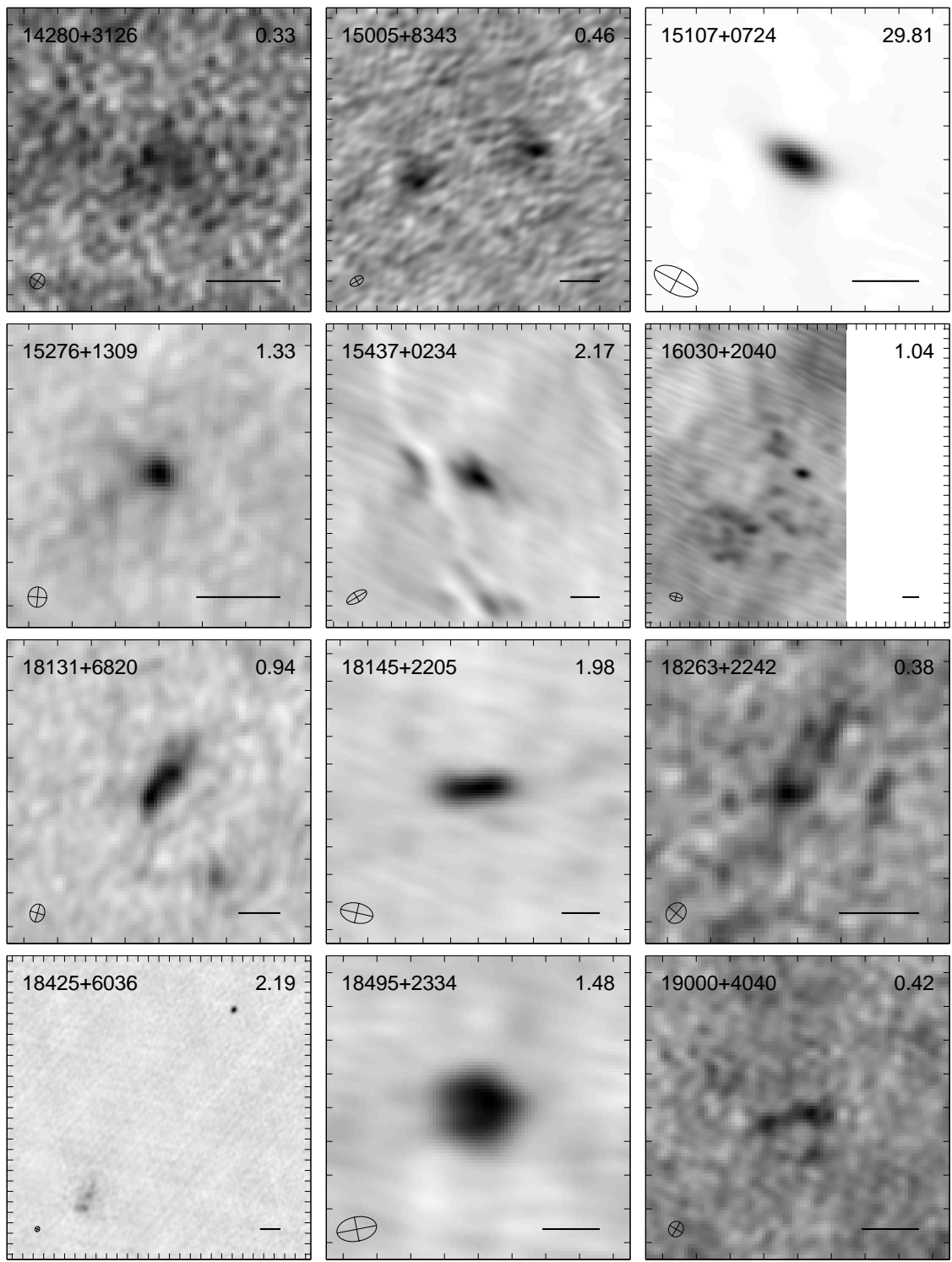}
\caption{See caption of Figure~\ref{fi:nondet1}}
\label{fi:nondet6}
\end{figure*}

\begin{figure*}[ht!]
\figurenum{3.7}
\centering
\includegraphics[width=0.9\hsize]{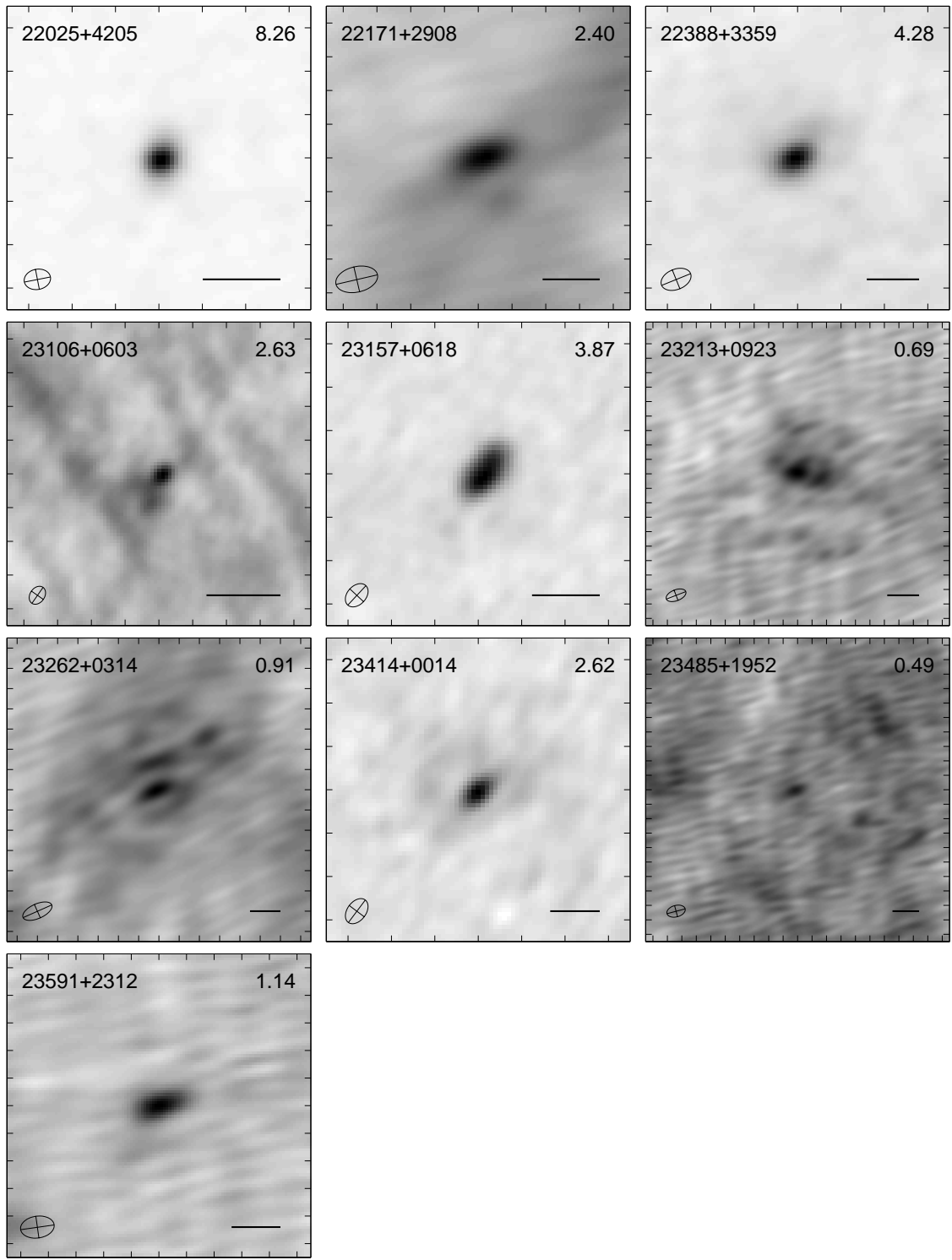}
\caption{See caption of Figure~\ref{fi:nondet1}}
\label{fi:nondet7}
\end{figure*}

\end{document}